\input harvmac
\settabs 3\columns \+&& SCIPP-03/06, RUNHETC-2003-16 \cr

 \vskip.4in \centerline{\bf A Critique of Pure String Theory: Heterodox Opinions of Diverse Dimensions } \vskip.3in

\centerline{\it T. Banks}\vskip.1in \centerline{Department of
Physics and Institute for Particle Physics} \centerline{University
of California, Santa Cruz, CA 95064} \centerline{and}
\centerline{Department of Physics and Astronomy, NHETC}
\centerline{Rutgers University, Piscataway, NJ 08540} \vskip.3in
\centerline{E-mail: \ banks@scipp.ucsc.edu} \vskip.5in

\centerline{\bf ABSTRACT}

\noindent I present a point of view about what M Theory is and how
it is related to the real world that departs in certain crucial
respects from conventional wisdom. I argue against the
possibility of a background independent formulation of the
theory, or of a Poincare invariant, Supersymmetry violating
vacuum state. A fundamental assumption is black hole dominance of
high energy physics.Much of this paper is a compilation of things
I have said elsewhere.  I review a crude argument for the
critical exponent connecting the gravitino mass and the
cosmological constant, and propose a framework for the finding a
quantum theory of de Sitter space.

\vfill\eject

\def\hn{{\cal H}_n}

\newsec{\bf Introduction: The Conventional Wisdom}

String theory, although it is a theory of gravity, is a creation
of particle physicists.  Traditional string phenomenology shows
its pedigree by asking for an exact solution of a purported theory
of everything, which exhibits exact Poincare symmetry (a symmetry
which is clearly only approximate in the real world).  This
theory is supposed to describe the scattering of particles in the
real world, which is thus postulated to be insensitive to the
cosmological nature of the universe.

The basis for this assumption is locality, a property that is
evidently only approximately true of string theory at low energy.
Super Planckian scattering is dominated by black hole
production\ref\tbfacv{R.~Penrose {\it unpublished} 1974; P.D.~
d'Eath, P.N.~ Payne,{\it Gravitational Radiation in High Speed
Black Hole Collisions, 1,2,3} Phys. Rev. D46, (1992), 658, 675,
694; D.~Amati, M.~Ciafaloni, G.~Veneziano, {\it Classical and
Quantum Gravity Effects from Planckian Energy Superstring
Collisions}, Int. J. Mod. Phys. A3, 1615, (1988), {\it Can
Space-Time Be Probed Below the String Size?}, Phys. Lett. B216,
41, (1989). H-J.~Matschull, {\it Black Hole Creation in
($2+1$)-Dimensions}, Class. Quant. Grav. 16, 1069, (1999);
T.~Banks, W.~Fischler, {\it A Model for High Energy Scattering in
Quantum Gravity}, hep-th/9906038; D.~Eardley, S.~Giddings, {\it
Classical Black Hole Production in High-Energy Collisions}, Phys.
Rev. D66, 044011, 2002, gr-qc/0201034. }\ , and the spectrum and
properties of black holes of sufficiently high energy are
definitely affected by the global structure of the universe.  By
continuity, there are effects on low energy physics as well.  The
only question is how large they are.

At any rate, a principal defect of this approach is that it
already postulates two mathematically consistent solutions of the
theory of everything, namely the real, cosmological, world, and
the exact Poincare invariant solution.  In fact, as is well known,
the situation is much worse than that.   There are many
disconnected continuous families of Poincare invariant solutions
of string theory.   They have various dimensions, low energy
fields, and topologies, but they all share the property of exact
SUSY.   The program of string phenomenology is to find a SUSY
violating, Poincare invariant solution of the theory, which
describes low energy scattering in the real world.  In
\ref\tbfolly{T.~Banks, {\it QuantuMechanics and CosMology}, Talk
given at the festschrift for L. Susskind, Stanford University,
May 2000; {\it Cosmological Breaking of Supersymmetry?}, Talk
Given at Strings 2000, Ann Arbor, MI, Int. J. Mod. Phys. A16,
910, (2001), hep-th/0007146} I expressed the opinion that no such
solution exists.   Be that as it may, the string phenomenologist,
having found the holy grail of a Poincare invariant, SUSY
violating, "realistic'' theory, will still be faced with the
question of why it is preferred over all of the vacuum states
with exact supersymmetry.

By contrast, if one adopts the hypothesis of cosmological SUSY
breaking (CSB) proposed in \tbfolly\ , this problem is solved at a
stroke. The theory of the real world has a finite number of
states\foot{The suggestion that the dS entropy represented a
bound on the number of states in dS space first arose in
conversations initiated by W. Fischler.  I asked Fischler to be a
coauthor on \tbfolly , but he declined on the grounds that he had
not contributed to the ideas about SUSY breaking.  Fischler
talked about the finite number of states at the Festschrift for G.
West in the Spring of 2000\ref\willy{W.~Fischler, {\it Taking de
Sitter Seriously}. Talk given at {\it The Role of Scaling Laws in
Physics and Biology (Celebrating the 60th Birthday of Geoffrey
West)}, Santa Fe Dec. 2000, and unpublished.}.} and can be neither
Poincare invariant, nor supersymmetric.  Since the number of
states in the real world is $e^{10^{120}}$, it would not be
surprising to find that ${\it some}$ of the properties of the
real world are well approximated by those of a Poincare invariant
theory, which I will call the {\it limiting vacuum}. By the
arguments (reviewed below) of \tbfolly\ , this limiting theory
must be SUSic and {\it have no moduli}. The combination of these
two properties and the general structure of SUSic theories imply
that it must be four dimensional, with only $N=1$ SUSY, and have
an exact complex R symmetry\foot{One must also use the fact that
the theory has a dS deformation to prove this.}. This puts strong
constraints on the low energy effective theory in the limiting
vacuum.  I have described approaches to the low energy
phenomenology of CSB in a recent paper\ref\scpheno{T.~Banks, {\it
The Phenomenology of Cosmological SUSY Breaking},
hep-ph/0203066.}.

In this paper I want to summarize a collection of ideas that I
have been playing with since 1999.  They form a context in which
the hypothesis of CSB is seen as a natural extension of the facts
we already know about M-theory.  These ideas are only loosely
connected and have not yet jelled into a consistent alternative to
the conventional wisdom about the way in which string theory is
connected to the real world.  I am setting them down here in the
hope that others can make more progress thinking about them than I
have.  If these ideas are even partially correct, then they imply
that some of our fundamental assumptions have been wrong, and I
think it is important that we revise them.

The key concepts revolve around the search for the fundamental
degrees of freedom of a quantum theory of gravity, and the
conviction that these are intimately connected with the high
energy behavior of the theory. All of our experience with quantum
theories suggests this connection.  This is nowhere more evident
than in Feynman's path integral formulation of quantum mechanics.
The key step in the derivation of the path integral is the exact
evaluation of the short time propagation kernel, and the key
assumption is that the short time behavior is dominated by a free
theory.  In this view, all of the formalism of classical mechanics
and canonical quantization is a consequence of the assumption of
what we have learned to call a Gaussian fixed point.  For this
reason, I will call theories whose short time behavior is
dominated by a Gaussian fixed point, Lagrangian theories.

Wilson's view of general quantum field theory as constructed from
relevant perturbations of general fixed point theories, may be
thought of as an extension of Feynman's principle. Again, the high
energy behavior defines the theory.  It has been our
fortune/misfortune to, for the most part, be able to access
non-Gaussian fixed point theories as infrared (IR) limits of
Lagrangian theories.  Another avenue to non-Gaussian fixed point
theories has been through cutoff models, mostly arising from
statistical mechanics.  The resemblance of the statistical sums in
these models to discretized path integrals has helped to obscure
the non-Lagrangian nature of the fixed points.  It is only with
the discovery of fixed points like the $(2,0)$ theory in six
dimensions, which have not yet been realized as infrared limits of
Gaussian models (this is impossible) or discretized statistical
sums (this is possible but unknown at the moment) that we have
been forced to face the truly radical departure from Lagrangian
dynamics that a non-Gaussian fixed point implies.

It is my opinion, that in attempting to construct a theory of
quantum gravity we should again look to the high energy behavior
of the theory.  When we do so, we are faced with several shocks.
Firstly, the traditional connection between high energy and short
distance disappears.  Even in perturbative string theory, high
energy physics is dominated by long strings.  More generally, in
any theory containing gravity there will be black holes.  The
Bekenstein-Hawking formula for the entropy of black holes suggests
that they dominate the high energy spectrum\ref\banksah{T.~Banks,
O.~Aharony, {\it Note on the Quantum Mechanics of M-Theory},JHEP
9903:016, (1999), hep-th/9812237. }, and semiclassical black hole
dynamics suggests that they are metastable.  Recent verifications
\ref\sssvetal{L.~Susskind, {\it Some Speculations About Black
Hole Entropy in String Theory}, hep-th/9309145; A.~Sen, {\it
Extremal Black Holes and Elementary String States},
Mod.Phys.Lett.A10:2081-2094,1995, hep-th/9504147 ; A.~Strominger,
C.~Vafa, {\it Microscopic Origin of the Bekenstein-Hawking Entropy
}, Phys.Lett.B379:99-104,1996, hep-th/9601029 .} that the
Bekenstein-Hawking formula indeed refers to a microscopic count
of all of the states of a black hole, lend credence to this point
of view.  Arguments to be reviewed below suggest that high energy
scattering processes are dominated by black hole production. The
result of these considerations is a radical new principle, which
I consider to be the ultimate form of the UV/IR correspondence :
{\it High Energy Dynamics is dominated by large black holes, some
of whose properties can be calculated using the semiclassical
Lagrangian formulation of general relativity.}  At the Davidfest
in Santa Barbara, I called this principle {\it Asymptotic
Darkness}.

The fact that certain features of black holes are describable in
the IR limiting theory is a direct consequence of the UV/IR
connection. The GR description of black holes is however
incomplete. It can give partial evidence for a huge set of states
associated with the black hole but cannot give a microscopic
quantum description of their properties.

The assumption that black holes dominate the high energy physics
of quantum gravity, and thus should be taken as a clue to the
whereabouts of the fundamental degrees of freedom, has several
dramatic consequences.  First of all, it immediately suggests the
Holographic Principle \ref\tHsuss{G.~'t Hooft, {\it Dimensional
Reduction in Quantum Gravity}, Essay published in Salamfest,
1993, p. 284, gr-qc/9310026 ;L.~Susskind, {\it The World as a
Hologram}, J.Math.Phys.36:6377-6396,1995, hep-th/9409089 } :
degrees of freedom should be associated with $(d-2)$ dimensional
areas in spacetime, rather than with points. At very high energy
densities, space is filled with black holes and the area scaling
of entropy becomes manifest. Fischler, Susskind and Bousso
\ref\FSB{W.~Fischler, L.~Susskind, {\it Holography and
Cosmology}, hep-th/9806039; R.~Bousso, {\it A Covariant Entropy
Conjecture }, JHEP 9907 (1999) 004, hep-th/9905177 ;{\it
Holography in General Space Times}, JHEP 9906 (1999) 028,
hep-th/9906022; {\it The Holographic Principle for General
Backgrounds},Class.Quant.Grav. 17 (2000) 997, hep-th/9911002.}
have shown how to formulate this principle for a general
spacetime. In spacetimes with appropriate\foot{I will explain
below why I think this idea does not generalize to the spacelike
boundaries of asymptotically de Sitter (dS) spaces.} asymptotic
boundaries one can see that this suggests a formulation in terms
of degrees of freedom on the boundary.  This fits in with
arguments from string theory and quantum gravity, that the only
observables in a theory of quantum gravity in asymptotically flat
or AdS spacetimes are boundary correlators like the S-matrix.

The AdS/CFT \ref\maldagkpw{J.~Maldacena, {\it The Large N Limit
of Superconformal Field Theories and Supergravity}, Adv. Theor.
Math. Phys. 2, 231, (1998), Int J. Theor. Phys. 38, (1999);
S.~Gubser, I.~Klebanov, A.~Polyakov, {\it Gauge Theory
Correlators From Non Critical String Theory } ,
Phys.Lett.B428:105-114,1998, hep-th/9802109 ;E.~Witten, {\it
Anti-De Sitter Space and Holography },
Adv.Theor.Math.Phys.2:253-291,1998, hep-th/9802150.}
correspondence is the most complete and successful realization of
this idea.  I want to emphasize that one can view the above line
of reasoning as a way of guessing or deriving the AdS/CFT
correspondence.  Namely, the spectrum of black holes in
asymptotically AdS spacetimes is that of a conformal field theory
living on the boundary.  Asymptotic Darkness,
 and the associated Holographic Principle then suggest
that the dynamics of the spacetime is completely captured by such
a conformal field theory (or a relevant perturbation of it).  We
now know that this is true in many cases.

For asymptotically flat spacetimes the consequences of the UV/IR
connection are more dramatic.  Black hole dominance implies that
quantum correlation functions of operators that do not distinguish
the degenerate microstates of black holes are not tempered
distributions in time, nor even the more singular distributions of
quasilocal field theories \banksah\ \ref\kap{A.~Kapustin, {\it On
the Universality Class of Little String Theories}, Phys. Rev.
D63, 086005, (2001), hep-th/9912044. }. This means there is no way
to localize the theory in time. Curiously, the black hole spectrum
{\it is} consistent with locality (quasilocality for four
asymptotically flat dimensions) in light cone time, which might
suggest a reason for the ubiquitous presence of the light cone
frame in Hamiltonian formulations of quantum gravity in
asymptotically flat spacetime.

A holographic formulation of nonperturbative quantum gravity in
certain asymptotically flat spacetimes is provided by Matrix
Theory \ref\bfss{T.~Banks, W.~Fischler, S.~Shenker, L.~Susskind
{\it M-theory as a Matrix Model: A
Conjecture},Phys.Rev.D55:5112-5128,1997, hep-th/9610043.}.  At
present it is formulated only in the approximation of discrete
Light Cone Quantization (DLCQ).

It is tempting to try to formulate a theory of asymptotically flat
quantum gravity in more covariant terms, as a theory on null
infinity.  Existing descriptions of massless particles at null
infinity suggest that one should not think of this as a dynamical
theory.  All the coordinates of null infinity are spatial (in the
sense that longitudinal and transverse coordinates are treated as
spatial coordinates in light cone frame).  Instead, dynamics is
encoded in the fact that null infinity is not a manifold , and
splits naturally into two disjoint cones with a common boundary at
spatial infinity.  Nontrivial correlation functions in the theory
are those which contain points on both components, and are nothing
but the matrix elements of the S-matrix.  The question that arises
(a question to which there is as yet no answer) is what the
dynamical principle is that determines the S-matrix. I will not
discuss null infinity much in this paper.

 The above discussion,
and \ref\isovac{T.~Banks, {\it On Isolated Vacua and Background
Independence} , hep-th/0011255.} make it clear (to me at least)
that the old dream of background independence in string theory is
a chimera. We already know that the various asymptotically AdS
spacetimes for which we have discovered the full quantum theory,
are not vacua of the same theory.  They are unitary quantum
theories without degenerate vacua \foot{Even the famous moduli
spaces are not there if we think of these as theories in AdS
space, so that the CFT is compactified on a sphere, rather than
as theories of branes embedded in flat space.} .  Some of them
are related by deformation by relevant or marginal {\it
parameters} or by compactifying one field theory and taking a
limit, but this is not what we usually mean by a theory having
multiple vacua.  It is also significant that the cosmological
constant in these theories is a discretely tunable, {\it
fundamental parameter} which encodes properties of the
fundamental UV theory, rather than a low energy effective
parameter, characteristic of a given IR representation of the
algebra of quantum operators (what we usually mean by a choice of
vacuum in QFT).

It is even more obvious that these are not vacua of a theory that
also includes asymptotically flat spacetimes, since the latter
have a radically different spectrum of high energy states, and
their state spaces carry representations of different maximal
spacetime symmetry groups.  Rather the two classes of theories are
related by the fact that the AdS theories are decoupling limits of
certain configurations in asymptotically flat space.  It is also
possible, that one can recover certain flat space theories by
taking large radius limits of AdS theories with free
parameters\ref\polchsuss{J.~Polchinski, {\it S Matrices from AdS
Space-Time}, hep-th/9901076; L.~Susskind, {\it Holography in the
Flat Space Limit}, Talk at the 8th Canadian Conference on General
Relativity, Published in Montreal 1999, General Relativity and
Relativistic Astrophysics, 98, hep-th/9901079. }. It is virtually
certain that not all asymptotically flat vacua can be retrieved
in this fashion.

In \isovac\ I argued that asymptotically flat theories also broke
up into disjoint families that are not states in the same
theory.  I will review these arguments below.

The key features unifying all of these bizarre properties of
quantum gravity are the fact that geometry responds to dynamics,
and the UV/IR connection, which intimately entwines the large
scale geometry (which in the traditional view is the vacuum
dependent part of the dynamics, to which the high energy behavior
is insensitive ) with the high energy spectrum (which I have
argued should be thought of as the domain where the fundamental
degrees of freedom are defined). This viewpoint suggests very
strongly that our traditional view of the cosmological constant
problem is in error.  The traditional view is that there is one
theory, which can have various infrared behaviors, characterized
by positive, negative or vanishing cosmological constant.  The
calculation of the cosmological constant in any given vacuum is a
dynamical problem: it is the calculation of the effective
potential for some low energy effective field.   Instead, the
UV/IR connection suggests that the cosmological constant is an
input, since it controls the very different behaviors of the high
energy density of states in the different theories.  This is
indeed true in AdS/CFT.  In known, supersymmetric, versions of
this correspondence, the cosmological constant is determined by
$N$, a parameter that characterizes the number of degrees of
freedom in the CFT. More generically, it will be completely
determined by the fixed point theory, even in those cases where
the full theory is a relevant deformation of the CFT.  The
cosmological constant is a property of the large scale,
asymptotically AdS geometry, which, by the UV/IR
correspondence\ref\sw{L.~Susskind, E.~Witten, {\it The
Holographic Bound in Anti-de Sitter Space}, hep-th/9805114.}, is
dual to the UV fixed point. The concept of an off-shell effective
potential, which combines information about theories with
different values of the cosmological constant, cannot be
meaningful if this point of view is correct.  At best it
corresponds to an approximate concept, valid only in extreme
regions of moduli space.

This line of thought leads inevitably to the conclusion of
\tbfolly\willy\ ,that the cosmological constant of an
asymptotically de Sitter (AsdS) space is a fundamental parameter
counting the number of states in the quantum theory.  I will
review the arguments for this below, and contrast this point of
view with that of \ref\edindia{E.~Witten, {\it Quantum Gravity in
de Sitter Space}, Talk given at Strings 2001, Mumbai, India,
hep-th/0106109 .} and \ref\dscft{A.~Strominger, {\it The dS/CFT
Correspondence}, JHEP 0110, 034, (2001), hep-th/0106113.} which
try to make a parallel with AdS/CFT involving the asymptotic
spacelike boundaries of AsdS spaces.   I will argue that the
latter approach neglects important back reaction corrections to
linearized classical gravity, and grossly overestimates the
number of observables in an AsdS space.  The key point is that
most of the measurements one can make on the (past) boundary
actually destroy the large scale geometry of dS space at a finite
time in the future (the CPT conjugate of this statement is also
true).

The final topic that I will discuss here is a similar sensitivity
of asymptotically flat geometry to the dynamics assumed for matter
inside it.  As of this date, we know of no example of a
controllable approximation to a theory of quantum gravity that
leads to a nonsupersymmetric theory in asymptotically flat,
Poincare invariant spacetime.  Within the regimes in which we are
able to calculate, we always generate a potential for moduli.  In
some cases we can argue that the moduli can be stabilized with
positive or negative cosmological constant.  In other cases, we
find Big Bang cosmological solutions with Newton's constant going
to zero in the asymptotic future.

This failure, and a perturbative string theory argument due to
Kutasov and Seiberg \ref\ks{D.~Kutasov, N.~Seiberg, {\it Number
of Degrees of Freedom, Density of States, and Tachyons in String
Theory and CFT}, Nucl. Phys. B358, 600, (1991).} lead me to
conjecture that there are {\it no} SUSY violating, Poincare
invariant theories of quantum gravity. Here again, I believe the
fundamental issue is the divergent spectrum of black hole states
in asymptotically flat geometries. SUSY cancellations are, I
believe, necessary for the existence of a sensible Poincare
invariant S-matrix with such a spectrum. This argument leads
directly to the conjecture\tbfolly\ that all SUSY breaking in the
real world should be associated with the existence of a positive
cosmological constant.  The latter conjecture has undergone a
fundamental change since I first suggested it.  In \tbfolly\ I
conjectured that it was virtual contributions of the largest
black hole states in dS space, which led to large
renormalizations of the classical formula for the gravitino
mass.  Psychologically, this had to do with my argument that it
was the black hole spectrum in asymptotically flat space that
required SUSic cancellations. I have since realized that most of
the states in dS space are not seen as black holes by any given
static observer. Rather, they are states on his horizon,
representing localized excitations seen by other observers. They
have static energy at most of order the dS temperature.  In the
limit in which dS space becomes Minkowski space, these low energy
states decouple from the Hilbert space of scattering states.  I
will argue below that the gravitino mass is due to its
interaction with these ultra-low energy states on the horizon,
which decouple from the limiting Poincare invariant theory.

The rest of this paper is an elaboration of the message of the
introduction.  Some of the arguments will be repeated in various
ways.  Feynman once said that if you have lots of arguments to
prove a given point, then that is a sign that you have no {\it
good} arguments.  I stand vulnerable to such an accusation.  I'm
writing this in the hope that someone smarter than I will pay
enough attention to these issues to come up with a good argument.

\newsec{\bf Supersymmetric Quantum Theories of Gravity}

I believe that the most cogent summary of what was achieved by the
String Duality Revolution is the statement that what we used to
call String Theory is really just the collection of Supersymmetric
Quantum Theories of Gravitation. Recall that using BPS arguments,
we can derive perturbative Type II string theories compactified on
tori as limits of the moduli space of a quantum theory, which has
another limit where its low energy dynamics is that of eleven
dimensional supergravity (SUGRA) (what some call M-theory but I
would prefer to call the 11D SUGRA limit of M-theory).

Similarly there is a collection of moduli spaces of theories with
16 supercharges, the most well known of which is 11D SUGRA on
$K3\times T^d$, with $d \leq 3$, also known as the heterotic
string compactified on tori.  Again, SUSY enables us to derive the
existence of this moduli space, including its weak coupling
stringy limits.  Perturbative string theory calculations give an
independent check that the SUSY arguments are valid.  In the
perturbative regime, we can of course do much more detailed
calculations of amplitudes.

The different moduli spaces discussed so far are disconnected,
and the arguments of \isovac\ , reviewed below, show that they
cannot be viewed as part of the same theory.   If we descend to 8
supercharges, it may be that the situation improves.   There are
many different descriptions of moduli spaces with 8 supercharges,
the simplest of which is Type II string theory compactified on
$CY_3$ manifolds.   The phenomenon of extremal
transitions\ref\gms{B.~Greene, D.~Morrison, A.~Strominger, {\it
Black Hole Condensation and the Unification of String Vacua},
Nucl. Phys. B451, 109, (1995), hep-th/9504145.} lends credence to
the conjecture that the moduli space of compactifications to four
asymptotically flat dimensions with 8 supercharges is a connected
space (though not a manifold). Furthermore, there exists the
possibility that we can recover all of the theories with more
supercharges by taking limits on this moduli space\foot{There is
a subtle issue here. We can certainly recover decompactified
theories with more SUSY by taking limits on this moduli space.
The question is whether we can actually take limits on the 8 SUSY
moduli space which give us the entire moduli spaces with more
SUSY. If we can only access these by decompactification and
recompactification then they are not really continuously
connected.}.

We also know how to obtain moduli spaces of SUSY theories with AdS
or linear dilaton asymptotics, as limits of brane configurations
in asymptotically flat theories.  These limiting theories are
quantum field theories and Little String Theories, respectively.
In these theories, the meaning of moduli space is somewhat
different\ref\seish{N.~Seiberg, S.~Shenker, {\it A Note on
Background (In)dependence}, Phys. Rev. D45, 4581, (1992),
hep-th/9201017.}. Fluctuating modes in these theories satisfy a
normalizability criterion at infinity, while changes in the
moduli correspond to non-normalizable perturbations.

In the case where the quantum theory is a QFT it is clear that
even changes in the moduli correspond to changes in the theory and
do not resemble the traditional notion of change of vacuum state
at all.  That is, they are simply changes in parameters specifying a
quantum field theory.
 Presumably, the same is true for the theories with
linear dilaton asymptotics.

In AdS/CFT we can find consistent relevant perturbations of the
CFT, which violate SUSY.  These have to be interpreted as quantum
theories of spacetimes which are asymptotically AdS but have local
distortions.  Note that these theories are not even Poincare
invariant, since in order to interpret them in terms of AdS
spacetime, we have to study them on the sphere.  But the full
superconformal algebra is restored asymptotically in spacetime.

Asymptotically AdS theories that break SUSY everywhere in
spacetime are harder to come by. We can, if we wish interpret any
conformal field theory as a theory of quantum gravity in
asymptotically AdS spacetime with radius of curvature of order
the Planck or string scale. Since there is no regime in which
ordinary low energy theories of gravity are valid in such a
spacetime, there is no calculation to compare the CFT results to,
which might contradict such an assertion.  The hypothesis that
there exist SUSY violating conformal field theories that can be
interpreted as AdS spacetimes with large radius leads to some
puzzles, which we will discuss below.

Apart from this possibility, we know of no examples of consistent
theories of quantum gravity in a large smooth spacetime, which
violate supersymmetry. If we try to break SUSY in perturbative
string theory, we generate potentials for moduli that lead either
to runaways or AdS vacua. The runaway solutions always drive the
system into regions where the perturbation series breaks down.
There is one class of models (the $O(16)\times O(16)$ heterotic
string in 10 dimensions is a typical example) where there are
solutions of the loop corrected equations of motion in which the
dilaton always stays in the weak coupling region although the
geometry has a Big Bang or Big Crunch singularity.   It is
possible that nonperturbative analysis can lead to consistent
theories of quantum gravity based on these solutions.

In my view, all of these facts are evidence that SUSY is a much
more crucial ingredient in theories of quantum gravity, than
semiclassical analysis has led us to suspect. This property of
quantum gravity is not at all evident in the field theory
approximation. One of string theory's most important contributions
to our understanding is to bring it to the fore.  I suggest that
we have not paid enough attention to what the theory is trying to
tell us. Followed to its logical conclusion, this clue resolves
many paradoxes and helps us to understand how string theory is
related to the real world.

I have left for last the discussion of supersymmetric theories of
gravity with Poincare invariance and only four supercharges.
These live in four or fewer spacetime dimensions.  It is likely
that there are infrared problems with the definition of
asymptotically flat theories of quantum gravity below four
dimensions, so I will restrict my attention to ${\cal N} =1$ SUSY
in four dimensions.  Here, the equations for SUSic, Poincare
invariant vacua, $W = D_i W = 0$ are over determined and we
expect few solutions, typically isolated.  It seems likely that
such isolated points have an enhanced discrete R symmetry that
accounts for the vanishing of the superpotential. Indeed, in the
presence of such an R symmetry it is sometimes natural to have
moduli spaces of ${\cal N} =1$ vacua\ref\bdmoduli{T.~Banks,
M.~Dine, {\it Quantum Moduli Spaces of $N=1$ String Theories},
Phys. Rev. D53, 5790, (1996), hep-th/9508071. }\foot{Witten has
pointed out another way to find moduli spaces of ${\cal N} =1$
vacua by using holomorphy and showing that the superpotential
vanishes order by order in the expansion about a weak coupling or
large radius region of moduli space. One must have control over
multi-instanton calculations to use this method.}. According to
the results of \isovac\ we should think of each of these
disconnected components of the four supercharge moduli space as
an alternative theory of gravitation in four asymptotically flat
dimensions. We will see that the isolated points in this moduli
space are of particular interest. None is yet known, because they
are not amenable to perturbative analysis. Indeed, we would hope
that they are few in number. Below, I will argue that the real
world is described by a quantum theory of gravity with a finite
number of states, which approaches an isolated minimal $4D$ Super
Poincare invariant theory in the limit that the number of states
goes to infinity. The limiting theory has to obey certain
additional constraints. If there is more than one such limit, we
would have to add experimental input beyond the existence of
gravitation and quantum mechanics to decide which of them is
related to the real world. The more isolated ${\cal N} = 1$ vacua
there are, the less predictive will be our theoretical framework.

In the following sections, I will try to attain a deeper
understanding of why SUSY is so important, and precisely what SUSY
is.

\newsec{\bf Classical Considerations or: Why General Relativity is Not a Field Theory}

One way to approach the quantum theory of gravity is to try to quantize the
classical Einstein equations.   Traditionally this has been done by treating
general relativity as a field theory.  Canonical quantization of a classical
system proceeds by finding a polarization of the symplectic structure on
its phase space; the space of solutions of the classical equations of motion.
The phase space of field theories is generally described by invoking the
Cauchy-Kowalevska theorem.  For hyperbolic theories in $d$ dimensions,
solutions are determined by fixing the field and its normal derivative on some
spacelike surface of dimension $d-1$.   The phase space is then said to consist
of one pair of canonical variables per space point.

While there does exist a Cauchy formulation of Einstein's equations, two facts
mitigate its utility.   The first is that the geometry of spatial surfaces
is time dependent and this (at least if we imagine imposing a fixed spatial
cutoff of short distances, like a simplicial decomposition of spatial
hypersurfaces) appears to present problems with unitarity.
The number of canonical degrees of freedom can change with time.

More problematic is the generic occurrence of singularities. The
Cauchy-Kowalewska theorem only guarantees local existence of
solutions. We have little rigorous data on the global structure of
the phase space of General Relativity.   What is known is that if
we parametrize solutions in terms of scattering data in
asymptotically flat space-time, or appropriate boundary data in
asymptotically AdS space-time (I will refer to both sorts of data
as scattering data from now on)  , then many singularities are
cloaked in black hole horizons.  Classically, this means that they
do not have any effect on the deterministic evolution of the
system outside the horizon\foot{Black Hole Complementarity is the
assertion that the same is true in the quantum theory, with {\it
deterministic} replaced by {\it unitary}.}.   A conservative
formulation of the Cosmic Censorship Conjecture is that this is
true of all solutions determined in terms of scattering data. I
will assume this form of Cosmic Censorship.

When combined with the Bekenstein-Hawking entropy formula for
black holes, this suggests that the field theoretic ``per unit
volume'' counting of degrees of freedom is wrong in General
Relativity.  Instead, as I will explain in more detail in the
next section, we have a holographic counting of degrees of
freedom\tHsuss\ .   Note this is also suggested by the idea that a
complete set of solutions should be parametrized by scattering
data.

Classical black holes are completely stable, and a specification of scattering data
would also have to list the number and positions of incoming and
outgoing black holes.
Quantum mechanically, because of Hawking evaporation,
particle scattering data are sufficient.

We see then that the attempt to quantize Einstein's equations semiclassically
leads us immediately into deep waters, and suggests a crucial role for Black
Holes in the fundamental formulation of the theory.  In the next section I
will argue that a more general approach to the quantum theory leads to the
same conclusion.

\newsec{\bf Short times at energy high}

The key step in the Feynman path integral formulation of quantum mechanics
is the approximation of the short time evolution operator by a perturbation
of a free system.  This leads to an expression for amplitudes as a path
integral over the exponential of the classical Lagrangian.  From these
formulae one can derive the standard canonical quantization procedure.
K. Wilson realized that this prescription did not always work in quantum field
theory, and needed to be generalized.  The Wilsonian definition of a quantum
system begins with a conformal fixed point theory and realizes a general
theory as a relevant perturbation of the fixed point.  The richness of
ordinary quantum mechanics and the relative scarcity of consistent quantum
field theories is a consequence of the fact that $1+0$ dimensional Gaussian
fixed points have a lot of relevant perturbations.

The above description of ordinary quantum mechanics and quantum field theory
highlights the fact that the fundamental description of the system
is obtained by looking at the high energy limit of its spectrum.
The ``degrees of freedom'' (DOF) of the system are a parametrization
of the high energy spectrum.  In models where the high energy behavior
is described by a Gaussian fixed point this coincides with the classical
definition of degrees of freedom.  For more general fixed points the
degrees of freedom have to do with the structure of the operator algebra.
In integrable CFT's there is usually a small set of generating operators
which can be thought of as the fundamental set of degrees of freedom.
At more general fixed points we do not yet know whether there is
such a simplification of the operator algebra.

{\it Asymptotic Darkness} is the conjecture that in all quantum
theories of gravity, the asymptotic spectrum of states is
dominated by black holes. There are several reasons for making
this conjecture.  The simplest is that the black hole spectrum
grows so rapidly at high energies and we have never discovered
another class of states with such rapid increase. More convincing
are arguments I will review below, which show that high energy
scattering processes produce black holes over a range of impact
parameters, which grows with the energy.  If these arguments are
correct, then all attempts to probe the theory at high energies,
probe black hole physics.  Finally, the AdS/CFT correspondence,
our most rigorously formulated quantum theory of gravity,
exhibits Asymptotic Darkness quite explicitly.  The form of the
high energy entropy, up to a multiplicative constant, follows
from symmetry arguments in the CFT and agrees with the Bekenstein
-Hawking entropy of AdS-Schwarzchild or Kerr black holes.  In the
one case where we can calculate the constant reliably from the
field theory, the answers also agree.

Asymptotic Darkness also sheds light on both the UV/IR connection
and the holographic principle.   If generic high energy states are
black holes of larger and larger mass, then we know that their
gravitational (and other) fields outside the horizon are smooth
and have low curvature.  Thus, there must be a description of some
of their properties in terms of long wavelength effective field
theory. Since the general laws of thermodynamics have to be obeyed
in the effective field theory approximation, it is not surprising
that semiclassical methods can be used to calculate the entropy of
black holes, even though the quantum states involved cannot be
treated correctly by effective field theory.  This, I believe, is
the ultimate form of the UV/IR connection.  Note that it also
implies that the form of the high energy spectrum is dependent on
the shape of spacetime at arbitrarily large distances.
Ultimately, it is the latter fact which is responsible for the
failure of the field theory paradigm of effective potentials and
superselection sectors.

The holographic principle follows from Asymptotic Darkness,
because the latter principle shows that the counting of high
energy states in the theory scales with the area, rather than the
volume, of the spacetime region they occupy.  In trying to probe
the theory to find the volume's worth of DOF we might have
expected, we would be forced to do scattering experiments at high
energies and small impact parameters.  Classical GR suggests
strongly that such experiments will always produce larger and
larger black holes, rather than probing short distances.

The intuitive argument for black hole creation in high energy
collisions goes as follows:  Imagine that a finite fraction, $M$,
of the energy of the collision remains for some time in a region
bounded by something of order the impact parameter.   Then we
have, at large distances from this region, a Schwarzchild field
with mass $M$.  For large $M$ and fixed impact parameter, the
Schwarzchild radius is larger than the region in which the energy
is concentrated and so the system must be a black hole. An
obvious loophole is the possibility that all but a finite amount
of the energy is radiated away.

The argument has been made more rigorous by a number of
developments\tbfacv\ . In $2+1$ dimensions, with negative
cosmological constant,
 the problem of colliding Aichelberg-Sexl waves
has been completely solved and black holes are indeed formed under
the stated conditions.  In $3+1$ dimensions, Penrose showed that
a trapped surface forms for collisions of zero impact
parameter.   If one invokes Cosmic Censorship, this shows that a
black hole forms and the size of the trapped surface puts a lower
bound on the mass of the black hole.  d'Eath and Payne studied
this problem in more detail. Eardley and Giddings have recently
generalized Penrose's argument to nonzero impact parameter. Thus,
at the level of classical general relativity the question of
black hole formation in such collisions has been reduced to the
proof of the Cosmic Censorship conjecture within the class of
solutions of Einstein's equations with scattering boundary
conditions.

It is important to realize that once this classical argument is
completed, we should believe that it is telling us something
correct about the quantum theory.  For large enough energy, the
fields outside the trapped surface are low curvature and
therefore well described by low energy classical field theory.
Again we see a manifestation of the UV/IR connection.

These classical arguments are not sufficient to tell us about the
detailed quantum mechanics of the final states of high energy
collisions. They do however tell us that it will be very
complicated.  High energy collisions at a range of impact
parameters from zero to an upper bound that grows like $E^{1\over
{D-3}}$ will be thermodynamic in nature, and the relevant
thermodynamics is that of black holes.

A final argument for black hole dominance in high energy
collisions comes from the AdS/CFT correspondence.  In this
context, black hole dominance just means thermalization.  There
is a lot of evidence that the high energy spectrum of states in
the relevant conformal field theories can indeed be identified
with AdS-Kerr black holes. Thus if we inject energy into the
system by making boundary perturbations, it follows from the
assumption that the dynamics of the CFT is not exactly
integrable, and the standard derivation (such as it is) of
statistical mechanics from quantum mechanics, that at
sufficiently high energy the system will thermalize.  The
identification of black holes with generic thermal states of the
system then shows us that high energy scattering leads to black
hole production.

To summarize, quantum theories are defined by their high energy behavior.
General arguments suggest that the slogan which captures the essence
of the high energy behavior of quantum theories of gravity is Asymptotic
Darkness.  We will see that this has important implications for understanding
how the various incarnations of M-theory are related to each other and to
the real world.

\newsec{\bf Against Independence}
I will use the term M-theory to refer to a collection of models of
the quantum theory of gravity, which we have been studying since
1984 (well, some of us (J. Schwarz) have been studying them since
1974) . They are for the most part SUSic, though there are some
SUSY violating systems that can be studied fairly reliably. An
important question is the extent to which these models are ``all
part of the same theory" , often called the question of background
independence. It is important to understand the precise meaning of
this, since any two separable Hilbert spaces are of the same
dimension are unitarily isomorphic to each other. So it is trivial
to map one theory onto another.  This is surely not what we mean
by background independence.

Our paradigm for what we do mean is classical field theory.  There
we have a Lagrangian density and different vacuum states of the
same theory mean different solutions of the same equations of
motion that preserve a maximal spacetime symmetry group.  In this
definition, we lump together Minkowski (M) space, Anti de Sitter
(AdS) space and de Sitter (dS) space, even though the meaning of
symmetry generators in the latter case is quite different. I will
argue that the conflation of these different kinds of spacetime is
incorrect in the quantum theory.

If we ignore the quantum mechanics of spacetime, and consider
quantum fluctuations in a fixed Minkowski space then there is a
nice quantum analog of this classical paradigm. Unitary Quantum
field theories are defined by relevant and marginal perturbations
of conformally invariant fixed point theories. Unitary Conformal
field theories have a unique conformally invariant vacuum state.
They are defined by a set of primary fields and their descendants
under the conformal group. This set of fields is in one to one
correspondence with the finite norm states of the theory. The
algebra of fields closes, in the sense that for every pair of
fields we have an operator product expansion (OPE): \eqn\ope{A(x)
B(0) = \sum C^n_i x^{d_n - d_A - d_B} O^n_i (0), } which converges
when applied to the vacuum state.

Some conformal field theories (usually SUSic ones) have a moduli
space of non conformally invariant vacua.  These are unitarily
inequivalent representations of the same local operator algebra.
The maximal symmetry of a state in these representations is the
Poincare subgroup of the conformal group.   These are examples of
what we mean quantum mechanically by different states of the same
theory.   The local operator algebra does not mix up these
different representations.   The complete Hilbert space of one
representation is obtained by taking limits of polynomials in
smeared local fields (with test functions of compact support)
acting on the vacuum. Often, the theories on the moduli space have
a particle interpretation.  That is, there are other bases for the
Hilbert space which consist of incoming and outgoing multiparticle
states.  The relation between the two descriptions of the Hilbert
space is given by the LSZ formula.  In this case the theory on the
moduli space always contains Goldstone bosons of spontaneously
broken scale symmetry.

Another way to get nonconformally invariant theories is to perturb
conformal field theories by a relevant operator.  This is the
Wilsonian definition of general quantum field theory.  Field
theories are parametrized by relevant perturbations of all
possible fixed points.   We do not normally think of field
theories with different values of their parameters as different
states of the same theory.  And indeed, there is a difference
between the breaking of conformal invariance along a moduli space,
and explicit breaking of conformal invariance by relevant
operators.  In the former case there is always a massless dilaton
in the theory, whose low energy couplings to other states is
characterized by conformal Ward identities.  So given a set of
Green functions which violate conformal invariance but approach a
conformal field theory at short distances we can tell whether they
represent a perturbed CFT or a moduli space of vacua by looking
for dilaton singularities.

Of course, nonconformally invariant field theories can also have
degenerate vacua. Again these are inequivalent representations of
the underlying operator algebra.  In generic CFT's the algebra is
harder to characterize.  Although it is in principle determined by
the first few terms in the short distance expansion, this
expansion is no longer convergent and the precise mathematical
characterization of what is going on is more difficult.  In QCD
this has led to endless questions about whether the perturbation
expansion determines the theory.  Nonetheless, the general picture
is clear.  Different states of a quantum field theory are
inequivalent representations of an underlying operator algebra
that can be extracted from a universal short distance behavior of
Green's functions.

There is a conceptually quite different way to discuss
inequivalent vacua in QFT.   Namely, given a single vacuum state,
one can, by injecting enough energy, construct regions of
arbitrary size that resemble another.  For vacua that are
continuously connected this is fairly trivial to do\foot{though
in theories of gravity it turns out to be surprisingly intricate
\isovac\ .}.  For isolated vacua it is a consequence of the
existence of static domain walls interpolating between two
vacua.  These are limits of large, long lived bubbles of one
vacuum, which can form inside another. I now want to discuss
whether either of these two methods of connecting vacua, works in
quantum theories of gravity.

\subsec{Disconnected asymptotically flat vacua}

In the last subsection, I reviewed two field theory methods for
judging when we have two vacuum states of the same theory. The
message of \isovac\ was that neither of these two methods of
verifying the existence of multiple states of the same theory
work in theories of quantum gravity unless there is a moduli space
of vacua.  Both fail because of the existence of black holes in
asymptotically flat space.  Quantum gravity appears to be a
holographic theory, which means that the only gauge invariant
observable in asymptotically flat space is the S-matrix.  The
closest analog of short distance behavior is the study of
scattering matrix elements in the limit that all kinematic
invariants are large.  In this regime the considerations of the
an earlier section lead us to expect scattering to be dominated by
the creation of supermassive black holes.  The Hawking
temperature of such holes is very low and thus the final
amplitudes are sensitive to the infrared structure of the theory
in its particular vacuum state.  There is no analog of the
univeral short distance behavior of different states of a QFT.

Similarly, a simple scaling argument shows that the attempt to
construct metastable bubbles of another vacuum generically leads
to the creation of a black hole.  Again the decay of the black
hole is sensitive to the nature of the external vacuum and
contains no trace of the putative information about the vacuum
state inside the black hole.

These arguments lead me to expect that isolated vacua of AF
quantum gravity are not ``states of the same theory" .   Something
similar can be said about disconnected pieces of moduli space. For
example, M-theory with 16 SUSYs has a disconnected moduli space
\ref\witdisc{E.~Witten, {\it Toroidal Compactification Without
Vector Structure}, JHEP 9802, 006, (1998), hep-th/9712028.}, one
branch of which is the conventional heterotic string on a torus.
The above arguments apply to the question of whether these
different branches are ``different states of the same
theory"\foot{This remark is correct within the context of moduli
spaces with 16 SUSYs.  If the maximal speculations about the
connectivity of the moduli space of 4D, 8 supercharge theories
are correct, and if we can obtain finite points in the 16 SUSY
moduli space by taking limits of the 8 SUSY moduli space, then we
could view all moduli spaces with 8 or more supercharges as one
connected theory.}.

How then are the various moduli spaces of vacua connected to each
other?   For vacua with at least $8$ supercharges there is a
plausible conjecture.   We have learned from Matrix Theory that,
in theories of quantum gravity, compactified spacetimes have more
degrees of freedom than their noncompact limits.  We also know
that many compactifications to four dimensions with eight
supercharges, lie on moduli spaces, and that many different
moduli spaces can be connected by extremal transitions.   It
would not contradict anything we know, to conjecture that there
is a single stratified manifold of four dimensional
compactifications with eight supercharges.  The quantum scattering
matrix\foot{Strictly speaking, there is no scattering matrix in
four dimensions, because of IR divergences.  However, there is a
generalized S-matrix with finite amounts of classical
gravitational radiation in initial and final states.} would vary
smoothly through the extremal transitions, as long as we kept all
stable states in the theory at all points in the moduli space. We
could further conjecture that all theories with at least $8$
supercharges and/or larger numbers of AF dimensions could be
achieved as limits along asymptotic directions in this moduli
space.  Many asymptotically AdS compactifications could also
arise as low energy limits of brane configurations in these
vacua.  This is, I believe, the maximal amount of ``background
independence'' that we can hope for in a quantum theory of
gravity.

The reader may wonder why I do not include theories with four
supercharges or fewer AF dimensions.  The latter are ruled out
because they do not really have S-matrices.  Indeed, although
there are perfectly good free string theories with 1-3 AF
dimensions, the vertex operator correlation functions that define
the perturbative S-matrix, do not exist. There is a good reason
for this.  Massive states, including most multiparticle states of
massless particles, distort the asymptotic geometry of spacetime,
so we can no longer talk about scattering theory in AF spacetime.
A hint of this problem already occurs in $4$ dimensions, as a
consequence of IR divergences.   Although a classical proposal
for a kind of scattering operator in $2+1$ dimensions
exists\ref\tHdj{S.~Deser, R.~Jackiw, G.~'t Hooft, {\it Three
Dimensional Einstein Gravity: Dynamics of Flat Space}, Annals
Phys. 152, 220, (1984).}, it is far from clear that there is a
good quantum theory. Furthermore, SUSic theories always contain
massless scalars as well as gravitons, so the mass of massive
states often appears infinite because of the contribution of
their long range scalar fields. Thus Super-Poincare invariant
theories of quantum gravity in $2+1$ dimensions may not even have
a collection of asymptotically locally flat scattering spacetimes
as in the proposal of \tHdj\ .

Four dimensional gravitational theories with $N=1$ SUSY do not
have a moduli space, at least generically. The equations $D_i W =
W = 0$ are overdetermined. They are likely to have solutions in
only two circumstances.  The first is at points of enhanced
discrete R symmetry. If some of the fields at such an enhanced
symmetry point have R charge zero, then we get a continuous
moduli space.  Otherwise, we expect only isolated vacua.  There
are also apparently\ref\edf{E.~Witten, {\it Non-Perturbative
Superpotentials in String Theory}, Nucl.Phys.B474:343-360,1996,
hep-th/9604030 } examples where instanton calculations and
holomorphy guarantee that the superpotential vanishes identically
along some subspace of a classical moduli space.

The disconnected moduli spaces of {\it e.g.} ten dimensional
theories with $16$ SUSYs may be connected through the moduli space
of theories with $8$ SUSYs in four dimensions. But there is no
apparent way to connect disconnected pieces of the moduli space of
four dimensional theories with minimal SUSY. We had previously
hoped to connect them by ``going over the potential barrier", but
the considerations of \isovac\ show that this does not happen.
Indeed, these considerations suggest that the notion of an
off-shell effective potential has no exact meaning in quantum
gravity, and is a useful tool only in extreme regions of moduli
space, if it is useful at all .

The latter remark is the most disturbing aspect of this critique
of the conventional wisdom.  So much of our effort has gone into
thinking about the calculation of the off-shell potential in
string theory that it is difficult to give up this tool.  Much
interesting recent work has gone into finding potentials which can
stabilize the moduli in regions where approximate methods of
calculation are valid\ref\evaetal{K.~Dasgupta, G.~Rajesh,
S.~Sethi, {\it M-Theory, Orientifolds and G-flux}, JHEP 9908,
023, (1999), hep-th/9908088; E.~Silverstein, {\it (A)dS
Backgrounds from Asymmetric Orientifolds}, hep-th/0106209;
S.~Giddings, S.~Kachru, J.~Polchinski, {\it Hierarchies from
Fluxes in String Compactifications}, Phys. Rev. D66, 106006,
(2002), hep-th/0105097. }. There do seem to be examples where the
quantum gravity effects I have discussed should not disturb the
existing calculations. However, the interpretation of the
potentials is far from straightforward, and deep issues of
quantum gravity in non-perturbative regimes are encountered in
this interpretation.   Another disturbing aspect of these
calculations is that the potential exhibits a plethora of AdS
vacua. Conventional wisdom would lead us to expect them to be
connected.  However, as the next section will show, this is
certainly untrue.

\subsec{The teachings of don juan: an AdS/CFT way of knowledge}

Our most comprehensive nonperturbative formulation of a quantum
theory of gravity is the AdS/CFT correspondence\maldagkpw\ . In
this subsection I want to explain how the ideas of asymptotic
darkness, inequivalence of vacuum states, and the nature of the
cosmological constant, appear in this formalism.

 The Bekenstein-Hawking formula for black hole entropy in AdS
spacetimes is identical with that of a conformal field theory.
This can in some ways be seen as a derivation of the AdS/CFT
correspondence, if we take asymptotic darkness as a fundamental
rule: black holes always dominate the high energy density of
states.  That is, accepting this rule, we derive from it the fact
that {\it quantum gravity in asymptotically AdS spacetimes is a
quantum theory which approaches a conformal field theory in the
UV} (at least with regard to its density of states).  But this is
nothing more nor less than the Wilsonian definition of a generic
quantum field theory.  The difference between general QFTs and
those which are actually conformally invariant has to do with the
rate at which AdS asymptotics are approached.  Thus, AdS/CFT is a
confirmation of asymptotic darkness, which is a more general
principle.

The AdS/CFT correspondence also throws light on the connectedness
of vacuum states.   In field theory, given two disconnected SUSic vacuum
states of the same theory, we expect a BPS domain wall interpolating
between them.  In fact, there are BPS domain wall solutions of
supergravity in situations where the AdS/CFT correspondence applies.
However, their meaning is radically different from what it was in
field theory.  The kind of BPS domain wall that has been discovered
in AdS/CFT is identified with {\it a
renormalization group flow between two different quantum field theories}.
Only one of these field theories has the full asymptotic density of
states of the system.  The other is a limiting infrared subspace of
the space of states.   Indeed, there is a c-function for this
flow, which decreases along it.  In the SUGRA approximation, decrease
of the c-function is a consequence of the same dominant energy
condition that leads to the black hole area theorem.

Strictly speaking, the BPS domain wall is not a part of quantum gravity
in an asymptotically AdS space at all.   The theory with AdS boundary
conditions is really the CFT on the sphere, and there is no global
BPS solution, because the domain wall has a translational invariant Poincare
energy density on a Poincare slice (and so is singular when mapped
to the sphere).  In the CFT this corresponds to the fact that, on
the sphere, RG flow is cut off in the IR by the finite volume.  Indeed,
AdS/CFT describes two related but distinct gravitational systems: globally
asymptotically AdS space, and the near horizon geometry of a BPS
brane in asymptotically flat space.   The BPS domain wall, like the
moduli spaces of vacua of the CFT, really exist only in the latter
interpretation of the theory.  However, the two forms of the theory
each contain complete information about how to construct the other, so
perhaps this is merely a technical quibble.

The important point is that,
although BPS domain walls exist in quantum gravity, they
no longer have the significance of connections between two vacua of
the same theory.  Rather they represent a theory with a dimensionful parameter,
 whose renormalization group flow interpolates between two
 conformal
(and therefore asymptotically AdS in the gravitational
interpretation) theories. However, it is only the UV fixed point,
which has the full set of degrees of freedom of the system, and
represents its true asymptotic behavior in spacetime.

It is important to note that the holographic RG flows are a
special kind of BPS domain wall, between a Breitenlohner-Freedman
allowed AdS maximum of the supergravity potential, and an AdS
minimum.  Other kinds of domain walls, including that between two
minima, do not have a C function which would enable us to
interpret them as RG flows (the C function flows backward for one
class of such walls, and is not monotonic for the other).  There
is no known dual description of such walls in CFT\foot{I would
like to thank M. Cvetic for teaching me about the different kinds
of BPS domain walls in AdS SUGRA.}. It seems entirely plausible
that they are a feature of low energy supergravities which does
not have a quantum mechanical realization in a complete theory.

There is an extension of these arguments which deals with the
question of whether AdS and Minkowski vacua are part of the same
theory. There are no known examples of a relevant perturbation of
a superconformal theory that could be identified with a domain
wall between an AdS and a Minkowski vacuum.   Let us recall the
geometric C-function that was discovered
\ref\gubseretal{D.Z.~Freedman, S.S.~Gubser, K.~Pilch,
N.P.~Warner, {\it Renormalization Group Flows from Holography,
Supersymmetry and a C Theorem}, Adv. Theor. Math. Phys. 3, 363,
(1999), hep-th/9904017.} in the parallel between the structure of
SUGRA domain walls and renormalization group flows
\ref\maldaetal{N.~Itzhaki, J.M.~Maldacena, J.~Sonnenschein,
S.~Yankielowicz, {\it Supergravity and the Large N Limit of
Theories with 16 Supercharges}, Phys. Rev. D58, 046004, (1998),
hep-th/9802042.}. The dominant energy condition gives a
directionality to the domain wall structure that mirrors the loss
of entropy as one flows from the UV to the IR in a field theory.
In domain walls interpolating between Minkowski and AdS vacua,
the C-function decreases from the Minkowski side. It is clear that
the Minkowski side of the domain wall is always in the UV and the
AdS side in the IR. Thus it is not surprising that one cannot find
domain walls that interpolate between Minkowski and AdS space as
relevant perturbations of a conformal field theory (for any such
theory is AdS invariant in the UV). AdS spaces should thus be
thought of as infrared limits of Minkowski space where a large
number of degrees of freedom decouple\foot{On the other hand, if
one has a sequence of AdS spaces with radii increasing to
infinity, it would not be surprising to recover an asymptotically
flat theory in the limit.  This is the approach taken in
\polchsuss\.  It should be noted that, although the arguments of
these authors are plausible, their construction is in no way a
complete argument for the existence of the unitary asymptotically
flat S-matrix that is their goal.  More importantly, it does not
seem likely that most asymptotically flat vacua of M-theory can
be recovered in this way.}.

The conclusion one is forced to accept, is that two different AdS
vacua are different quantum systems, rather than two states of the
same system. The clearest way of explaining the disconnectedness
of AdS vacua, is to note that the cosmological constant in Planck
units is,{\it in every AdS/CFT correspondence of which I am aware,
a discrete tunable variable, which characterizes different field
theories ({\it e.g.} a power of the $N$ of an $SU(N)$ gauge
group).} Nothing could be further from the idea of the value of an
effective potential of a given theory at its different minima. The
effective potential is a strictly IR concept, and different ground
states of a field theory have the same UV behavior.  By contrast,
in AdS/CFT, different values of the cosmological constant
correspond to different high energy behaviors, since $N$ controls
the density of states at high energy.  These results are in accord
with (in fact were the origin of) the conjecture of Asymptotic
Darkness. The generic high energy state in AdS space is a black
hole, and the nature of the black hole spectrum is crucially
dependent on the cosmological constant, because the entropy of a
black hole is determined by the area of its horizon.

I conclude from this that the field theoretic notions of off shell effective
potential and the possibility of connecting different vacuum states `` through
the off shell configuration space'', are not valid concepts in theories
of quantum gravity in asymptotically AdS space.  The infrared properties
of spacetime, which we generally associate with the nature of the vacuum
state, are determined by the UV properties of the quantum field theory
which defines quantum gravity in asymptotically AdS spaces.

Since the quantum theory of gravity in asymptotically AdS spaces
is defined by a quantum field theory, it is interesting to ask
what the interpretation of QFT superselection sectors, multiple
vacuum states, and effective potentials are in the spacetime
interpretation.  The key to this question is the proper
formulation of asymptotically $AdS_d$ gravity as QFT quantized on
$S^{d-2} \times R$.  We have already seen that (if $d > 3$) this
eliminates the moduli spaces of CFT's .   However, it might appear
that we could still get multiple vacua, by considering relevant
perturbations of the CFT that lead to a theory with a field
theoretic (FT) effective potential with multiple minima. Relevant
perturbations of a CFT lead to a quantum theory of spacetimes
which asymptotically approach AdS but not quite fast enough that
the perturbation can be considered a normalizable fluctuation.
They seem to be perfectly legitimate quantum theories of gravity,
which describe inhomogeneous spacetimes with only spherical
symmetry and a single global timelike Killing vector.  All other
generators of the conformal group have been broken by the
combination of the relevant perturbations and the asymptotically
AdS boundary conditions. Perhaps general renormalizable QFTs
should be thought of as theories of defects with long range
gravitational fields, embedded in an AdS spacetime.

When renormalizable QFTs are quantized in Minkowski space, they
can have multiple vacua. Once again, however, the fact that we are
quantizing on a compact space defeats our attempt to find multiple
vacuum states. Discrete degeneracies of the classical FT potential
on a compact space, lead to a unique ground state, which in the
the semiclassical approximation is a superposition of
semiclassical ground states in each of the degenerate wells.  At
large AdS radius, the quantization sphere is large, so the
tunneling amplitudes between semiclassical ground states are
small, of order $e^{-(mR_{AdS})^{d-2}}$, where $m$ is the mass
scale induced by the relevant perturbation. However, the large
distance behavior {\it in space time} is UV behavior in the field
theory and is insensitive to the choice of ground state. Thus,
although we could set up long lived states which live close to one
minimum of the FT effective potential, the asymptotic observer
will not view them as different vacua.   Similarly, if we
construct a FT effective potential with a metastable minimum, we
will indeed find a metastable state in the quantum theory, but its
decay will look nothing like a Coleman
DeLuccia\ref\cdl{S.~Coleman, F. De Luccia, {\it Gravitational
Effects On and Of Vacuum Decay}, Phys. Rev. D21, 3305, (1980).}
bubble to the asymptotic observer. Rather, it will look like the
decay of a localized object, which does not affect the asymptotic
structure of spacetime.

I find these translations of the known field theoretic structures
that we have been trying to mimic in thinking about vacuum states
of quantum gravity, to be the most convincing evidence that we
have been deluding ourselves.  The notions of vacuum, effective
potential, and vacuum decay from field theory, are not correct
ones in quantum gravity\foot{As discussed in
\ref\heretic{T.~Banks, {\it Heretics of the False Vacuum:
Gravitational Effects On and Of Vacuum Decay, 2.},
hep-th/0211160.}, this might not be true in de Sitter quantum
gravity. As that theory has not yet been constructed, it is
harder to assess the validity of these concepts in a rigorous
manner. There seems to be a sensible semiclassical theory of dS
decay into other dS spacetimes or into negatively curved FRW
universes with vanishing cosmological constant. }
 Similar remarks could have been made about asymptotically flat vacua
in the context
of Matrix Theory, but the problem of seeing phenomena associated with
the vacuum in light cone quantum field theory, made one suspicious of
the conclusions.

I want to end this subsection with a few remarks about SUSY
breaking in AdS spaces.   One way to do this is to add relevant
SUSY violating operators to a known large radius AdS/CFT
correspondence.  This is not terribly interesting.  As we take
the AdS radius to infinity, we find that most of the spacetime is
SUSic.  We are really looking at some sort of SUSY violating
defect in a SUSic vacuum state.  It is more interesting to study
SUSY violating fixed point theories.  Here the problem is to find
large radius examples.   An interesting class of examples is
given by the large $N$ version of the fixed points studied in
\ref\bankszaks{T.~Banks, A.~Zaks, {\it On the Phase Structure of
Vector-Like Gauge Theories With Massless Fermions}, Nucl. Phys.
B196, 189, (1982). }. These are large N gauge theories with
massless fermions in $N_f$ copies of the fundamental. $N$ is
taken to $\infty$ with $N/N_f$ fixed in such a way that the
$\beta$ function for the gauge coupling has a zero in the
perturbative regime $g^2 N \sim \epsilon \ll 1$.  One expects, on
grounds of continuity, and because of the known examples with
${\cal N} = 1$ SUSY, that there is a collection of such fixed
points with $\epsilon$ ranging from a very small number to
something of order $1$.   These theories have a $1/N$ expansion,
whose leading term consists of planar diagrams with an arbitary
number of fermion loops (holes).   The planar limit is {\it not}
a free string theory, because there is no restriction of
$SU(N_f)\times SU(N_f)\times U(1)$ gauge invariance. The only
obvious conjecture to make about what the large $N$ expansion
means in a dual spacetime picture is that it is the low energy
SUGRA expansion. This would be remarkable if true because it
would mean that we could solve a rather nontrivial four
dimensional CFT by solving Einstein's equations. However, because
the theory has a huge number of operators whose dimension is a
finite multiple (as $N\rightarrow\infty$) of that of the stress
tensor, it seems that, as in perturbative string theory with AdS
radius of order the string scale, there is no regime in which
SUGRA is a good approximation.

A putative AdS dual of these field theories would have some
peculiar properties.  The global $SU(N_f)\times SU(N_f)\times
U(1)$ symmetry of the boundary CFT implies a huge spacetime gauge
group in the limit of large AdS radius. The simplest context for
investigating this phenomenon is probably the ${\cal N} = 2$
superconformal analogs of these non-SUSic fixed points.  A
possible interpretation of the huge spacetime gauge group is the
following:  $A_n$ singularities in string theory can produce
$SU(n+1)$ gauge groups.  D-branes at the weakly coupled string
orbifold singularity of $A_n$ type give rise to a ${\cal N} =2 $
quiver gauge theory\ref\mrdgm{M.R.~Douglas, G.~Moore, {\it
D-Branes, Quivers, and ALE Instantons}, hep-th/9603167.}. The
string theory orbifold differs from the configuration with the
true $A_n$ singularity by having NS B fields on the shrunken
cycles, but the B field can be continuously dialed to zero.   One
then conjectures that the superconformal theory with $N_f = 2
N_C$ can be achieved as a marginal perturbation of the quiver
theory, and that its spacetime interpretation is that of $N$
branes at an $A_{2N-1}$ singularity, without B flux\foot{This
conjecture was developed with B. Acharya and H. Liu.}.

One can obtain non-SUSic fixed points without flavor groups by
studying large $N$ theories with a variety of fermion
representations, tuned so that the leading coefficient of the
Callan-Symanzik function is small.  Little is known about such
theories, but the AdS/CFT correspondence suggests that they have
an  AdS dual with radius large compared to the Planck scale.

There is however a general problem with all of these constructions
of SUSY violating, large radius AdS space.  By making $N$ large,
we guarantee that the AdS radius is large compared to the Planck
scale (comparing the Bekenstein-Hawking entropy formula with the
CFT entropy formula).   But there is no guarantee that it is
large compared to the string scale.   All of these constructions
are reminiscent of large $N$ gauge theories with fixed 't Hooft
coupling.  Such theories have a large number of operators (going
to infinity with $N$) whose anomalous dimension is a finite
multiple of that of the stress tensor.  Thus, as in string theory
with weak string coupling, there is a spectrum of particles whose
mass is of order the inverse AdS radius, in the limit that the
Planck mass goes to infinity.  Furthermore, E.Gorbatov\foot{E.
Gorbatov, private communication} has argued using the
Horowitz-Polchinski correspondence principle, that in this range
of couplings there will be no black hole states with size less
than the AdS radius.   The large $N$ limit of these theories is a
free theory of an infinite number of free massive particles in
AdS space, rather than an interacting theory of quantum gravity
in flat space.   This is consistent with our conjecture in a
later section, that SUSY violating theories of gravity in
asymptotically flat space do not exist.

Special mention should be made here of the SUSY violating
orbifolds of $AdS_5 \times S^5$ gravity\ref\ksnv{A.E.~Lawrence,
N.~Nekrasov, C.~Vafa, {\it On conformal Field Theories in Four
Dimensions}, Nucl. Phys. B533, 199, (1998), hep-th/9803015;
S.~Kachru, E.~Silverstein, {\it 4-D Conformal Theories and
Strings on Orbifolds}, Phys. Rev. Lett. 80, 4855, (1998),
hep-th/9802183.}. These are described by non-supersymmetric
quantum field theories, which nevertheless have a line of fixed
points in the planar limit. The gap in dimensions between the
stress tensor and (most) other operators, which we expect in a
theory which describes a space with curvature low compared to the
string scale, is guaranteed by the analogous phenomenon in the
SUSic parent theory.

However, like all tree level SUSY breaking in string theory, this
example is highly unstable.  Non-planar corrections give infinite
mass to the scalars and the theory nominally flows to an
asymptotically free theory of gauge bosons and fermions.  If we
try to tune the scalar masses to zero we encounter the
Halperin-Lubensky-Ma-Coleman-Weinberg\ref\hlmcw{B.~Halperin,
T.~Lubensky, S.K.~Ma, {\it First Order Phase Transitions in
Superconductors and Smectic A Liquid Crystals}, Phys. Rev. Lett.
32, 292, (1974); S.~Coleman, E.~Weinberg, {\it Radiative
Corrections as the Origin of Spontaneous Symmetry Breaking},
Phys. Rev. D7, 1888, (1973).} first order fluctuation induced
phase transition, rather than a conformally invariant fixed
point. Even if we assume that some conformal point can be found
non-perturbatively, the nonplanar corrections to the 't Hooft
coupling's Callan-Symanzik function show that any such conformal
theory will be an isolated fixed point with $g^2 N$ of order
$1$.  In other words, these orbifold theories behave in a manner
similar to Scherk-Schwarz orbifolds of toroidal string theories.
At tree level they give a SUSY violating string spectrum in a
large smooth spacetime with a maximally symmetric subspace.  Once
loop corrections are taken into account this picture is not even
approximately valid.  As a field theory the theory likely becomes
asymptotically free in the UV and has no large smooth spacetime
interpretation.   It is possible that there is a conformal point
that can be defined as a limit of the cutoff version of this
field theory, but it can at best be interpreted as an AdS
spacetime with radius of order string scale.

Finally, I would like to mention SUSY violating flux
compactifications of string theory, which can describe large
radius AdS space-times for appropriate values of flux \evaetal\ .
There is no contradiction between these models and the statements
I have made here, because the SUSY violation goes away as the
radius is taken to infinity.   Nonetheless, they are peculiar
from the point of view of AdS/CFT.  Since many of them are
effectively perturbative, we might expect them to be related to
large $N$ gauge theories in three dimensions. Thus, for large
values of the flux, these perturbative vacua predict infinite
sequences of fixed points at large $g^2 N$ in some kind of large
$N$ gauge theory.   It would be of great interest to get a handle
on these peculiar conformal field theories. Alternatively,
evidence that they did not exist would throw doubt on the
effective potential calculations that went into the effective
field theory constructions of these systems.

 \subsec{Flat contradictions}

The ordering of Minkowski with respect to AdS in our discussion
of domain walls, fits with a striking difference between the black
hole entropy formulae for the two spacetimes. Indeed, using the
logic of the previous subsection, it is clear that quantum
gravity in asymptotically flat spacetimes is a different kind of
beast from quantum field theory, with a high energy density of
states unlike any quantum theory we have dealt with before.  It
grows more rapidly than an exponential of the energy.   This
means that Green functions of generic Heisenberg operators:
\eqn\gfcn{<0| O(t_1 ) \ldots O(t_n ) |0>} are not tempered
distributions\banksah\kap\ . That is, if we smear the operators
with smooth functions of compact support, they are not finite.
Consequently, we should not expect to be able to localize
operators in time\foot{The exceptions would be operators that
resolve the degeneracy of black hole states and have matrix
elements to only a few states of arbitrarily high energy. Given
the thermal nature of black hole decay, it is likely that the
description of such operators in terms of the asymptotic particle
basis is hopelessly complicated.}.  Much of the conventional
framework of quantum mechanics is lost.

This new feature of asymptotically flat space is also suggested by
the structure of the boundary of Minkowski space, and the
description of physics on it. Ashtekar\ref\asyquant{A.~Ashtekar,
{\it Asymptotic Quantization}, Naples Italy, Bibliopolis (1987),
Monographs and Textbooks in Physical Science, 2.} has described
the quantum theory of massless particles on null infinity.  In $d$
dimensional AF spacetimes, future (or past) null infinity is a
$d-1$ manifold with a degenerate conformal structure.  There
exists a coordinate system $(u, \Omega)$ where $u$ is null and
$\Omega$ coordinatizes a manifold with metric conformal to the
round $d-2$ sphere. Ashtekar uses Fock space methods to define
multiple asymptotic massless particle states in terms of fields
defined on null infinity.  The commutation relations of the
fields have the form \eqn\comrel{[F(u,\Omega ) , F(u' ,
\Omega^{\prime} ) ] =3D \Delta (u - u^{\prime}, \Omega,
\Omega^{\prime} )}. Thus, all of the coordinates of null infinity
are spatial.  The fields do not solve any dynamical equations on
null infinity, but are independent variables at each point. Thus,
we should not expect a parallel to the AdS paradigm of
"correlation functions on the boundary", for asymptotically flat
spaces. Instead, the dynamics will be incorporated in a mapping
(the S-matrix) between formulations of the theory on past and
future null infinity. That is, the boundary of asymptotically
flat space splits into two pieces, neither of which is compact.
They are joined along an asymptotic region (spacelike infinity)
that they both share.  The Hilbert spaces associated with the two
boundaries carry unitarily equivalent representations of the
Poincare group, but there is a nontrivial Poincare invariant
unitary operator, the S-matrix, which connects them.  What we do
not yet have is an exact prescription for calculating that
S-matrix.

Perturbative string theory describes it as an asymptotic series
in a small parameter.   Matrix theory \bfss\ describes it as the
(conjectural, Poincare invariant) large $N$ limit of the S-matrix
of an $N\times N$ matrix quantum mechanics. Polchinski and
Susskind \polchsuss\ have suggested a way to obtain it as a limit
of CFT correlation functions. Aside from questions of
convergence, it is clear that none of these prescriptions applies
to all situations in which we expect to have Poincare invariant
vacua of M-theory.

Another approach to a holographic description of asymptotically
flat spacetime is to use the light-front gauge, as in
perturbative string theory and Matrix theory.   The black hole
spectrum throws light on the ubiquity of light-front gauge in
Hamiltonian descriptions of asymptotically flat M-theory (AFM).
Indeed, for $5$ or more asymptotically flat directions, the light
cone energy spectrum (equivalently, the $M^2$ spectrum for fixed
longitudinal and transverse momentum) grows more slowly than an
exponential so that conventional quantum mechanical formulae make
sense in light cone time.  It is quite interesting that this
argument (marginally) fails in four dimensions, where asymptotic
darkness predicts a Hagedorn-like exponential spectrum.  Perhaps
this means that the Hamiltonian theory of four dimensional
asymptotically flat spacetime is a quasi-local field theory, or
little string theory, in light cone time.

The special role of four dimensions appears in a number of other
contexts. It is the lowest dimension in which asymptotically flat
spacetime has black hole excitations, indeed the lowest dimension
in which it has any massive excitations at all (and remember that
generic multiparticle states of massless particles are massive).
It is also the dimension where the S-matrix ceases to exist in
any quantum theory of gravity.  Infrared divergences cause the
vanishing of all amplitudes, which do not have an infinite number
of gravitons in the final state.   As a consequence, one must
invent a generalized S-matrix between states with coherent
classical gravitational radiation. The asymptotic symmetry group
of this class of spacetimes is the Bondi- Metzner-Sachs (BMS)
group and the theory is undoubtedly more complicated\foot{In
fact, the relevant Lie algebra may be even more interesting than
that of BMS.  Let $(u,\Omega)$ be the coordinates of (say) future
null infinity, $u$ a null coordinate and $\Omega$ a coordinate on
the $d-2$ sphere. The BMS algebra is the semidirect product of
vector fields of the form $f(\Omega )\partial_u$, with $f$ an
arbitrary function on the sphere, and the conformal algebra of
the sphere (with conformal factor also rescaling $u$).  For $d=4$
the conformal algebra is the infinite dimensional Virasoro
algebra. The correct symmetry algebra of the formalism is
probably this large extension of the BMS algebra, or some
interesting subalgebra of it {\it e.g.} $f$ might be restricted
to be the sum of a (locally) holomorphic and anti-holomorphic
function on the sphere.} than it is in higher dimensions.

At the present time, we do not have a general prescription for
writing down the quantum theory for asymptotically flat
spacetimes.  For spacetimes with six or more asymptotically flat
dimensions and 16 or 32 supercharges, Matrix Theory provides a
plausible answer, though of course one has yet to prove Poincare
invariance of the large $N$ limit.  Matrix Theory is supposed to
be the Discrete Light Cone Quantization (DLCQ) of M-theory. The
spectrum of the DLCQ theory diverges more rapidly at large energy
than that of the limiting, decompactified theory, for $9$ or
fewer asymptotically flat dimensions.  At $D=5$ \foot{at least
for toroidal compactification.  The authors of
\ref\KLS{S.~Kachru, A.~Lawrence, E.~Silverstein, {\it On the
Matrix Description of Calabi-Yau Manifolds}, Phys. Rev. Lett. 80,
2996, (1998), hep-th/9712223. } claim that for Calabi-Yau
compactification the DLCQ is some kind of $3+1$ dimensional field
theory.} the DLCQ spectrum blows up faster than an exponential of
light cone energy and we don't know how to define it.  It is of
the greatest interest to work out the form of the decompactified
quantum theory.

Another approach to flat spacetime is to take the large radius
limit of AdS/CFT.  This applies to even fewer examples, and the
construction is on a much less firm footing.  In Matrix Theory,
one has a well defined, unitary scattering matrix and one must
show that its limit exists and is Poincare invariant.  In the
AdS/CFT approach one has a complicated definition of scattering
matrix elements, and one must prove that they form a unitary
matrix as well as proving that the proper flat space symmetry
group (which is larger than the contraction of the AdS/CFT
symmetry group) is restored.

\newsec{\bf The Peculiar Position of Perturbative String Theory}

In my description of M-theory in the second section, strings were
exiled to certain extreme regions of moduli space.  Perturbative string
theory was useful for confirming dualities, and for doing certain
exact calculations, which are protected by SUSY and have a nonperturbative
significance.

It is clear however that there is more to perturbative string
theory than that.  It contains baby versions of the holographic
principle, the UV/IR connection, the growth of the high energy
spectrum of states, all of which we have seen to be central
features of the quantum theory of gravity. String theory also
teaches us about the importance of SUSY in the theory. There are
no known asymptotically flat string vacua with broken SUSY.
Kutasov and Seiberg\ks\ have given a very general argument for why
this is so.

Historically, perturbative heterotic string theory gave us the
first indication that a theory of gravity could also explain the
standard model of particle physics.  The more recent and more
general understanding of how non-abelian gauge groups and chiral
fermions arise from singular limits of Kaluza-Klein SUGRA does
not diminish the historical importance of the perturbative string
results.  Indeed, the geometric picture was derived
\ref\hulltowns{C.M.~Hull, P.~Townsend, {\it Unity of Superstring
Dualities}, Nucl. Phys. B438, 109, (1995), hep-th/9410167.} by
trying to understand how to reproduce the heterotic string
results in the dual Type II picture.

It is even more remarkable that, in a variety of situations, perturbative
string theory and duality have allowed us to guess/derive the exact
non-perturbative formulation of the theory.  We would have neither Matrix
Theory nor the AdS/CFT correspondence if it were not for the tools of
perturbative string theory.

These facts suggest a deeper role for perturbative string theory
than I have allowed it in these ruminations.  What it might be is
beyond my comprehension.  It seems unlikely to me that the answer
is String Field Theory. Witten's formulation of classical open
bosonic string field theory is undoubtedly
elegant\ref\wsft{E.~Witten, {\it Noncommutative Geometry and
String Field Theory}, Nucl. Phys. B268, 253, (1986).}. But this
classical theory misses a lot of gravitational physics that seems
essential.  In particular, it makes no distinctions between
phenomena on D-branes of different co-dimension. It is clear that
once gravitational interactions come into play, the corrections to
classical open string field theory for D-branes of low
codimension are large (generally infinite). More generally, the
loop effects which give rise to gravitons in open string
perturbation theory are singular.   If the graphs are regularized
in any known way, one finds that one must add a divergent series
of corrections to the Lagrangian to reproduce perturbation theory
to all orders (generally one has to add an explicit closed string
field as well).  So String Field theory does not give us a
non-perturbative definition of a quantum theory.

It seems more likely to me that the elegant connection between
classical string theory and the world sheet renormalization group
is the place to search for a deeper connection between the
perturbative string formalism and a well defined non-perturbative
theory of quantum gravity.

One avenue of research which, as a consequence of the
considerations of this paper, seems very unlikely to lead to
successful results, is the search for a perturbative string
theory resolution of space-like cosmological or black hole
singularities.   This search was based on the notion that string
theory was the correct high energy description of theories of
quantum gravity.   The breakdown of perturbative string theory in
a wide range of the kinematic regime that we can call high energy
(including both the tradition Regge and fixed angle regimes)
suggests very strongly that it is not.

\newsec{\bf de Sitter Space: the importance of being finite}

Asymptotically dS spaces, according to the conjecture of
\tbfolly\willy\ , are described quantum mechanically by finite
dimensional Hilbert spaces.  This fits in well with what we have
learned from the AdS/CFT correspondence.  We have learned that
negative values of the cosmological constant are discrete
parameters (partially) determining different theories of quantum
gravity, and controlling the high energy density of states in the
theory.  Similarly, the finite Hilbert space conjecture identifies
positive values of the cosmological constant as a discrete
parameter, wholly or partially determining different theories of
quantum gravity.  In this case the cosmological constant
determines an upper cutoff on the energy spectrum of the theory
(in static coordinates), equal to the mass of the Nariai black
hole.

A question that arises immediately is where representations of the
dS group fit into such a story. There are several answers to this
question, depending on what one is trying to accomplish. The
simplest\edindia\ is to claim that since the global dS spacetime
has no spatial boundary, all observables, and physical states are
dS invariant. However, this does not take into account the fact
that observation in physics always consists in separating the
world into an experimental apparatus, and a system.  A realistic
measuring apparatus will follow a timelike trajectory in dS space
and determine a static coordinate system. The cosmological
horizon volume of this static system is, in $dS_d$, invariant
under an $R \times SO(d-1)$ subgroup of $SO(1,d)$. $R^+$ is
generated by the static Hamiltonian.  This subgroup will act as a
group of global symmetries on the quantum mechanics of this
observer.  Thus, by a choice of gauge, we introduce a boundary
into the system, on which to define those generators of the dS
group which preserve the gauge as global symmetry operators.
Gomberoff and Teitelboim\ref\gt{A.~Gomberoff, C.~Teitelboim, {\it
De Sitter Black Holes With Either of Two Horizons as a Boundary},
hep-th/0302204. } have given a rather explicit description of how
this works for all Kerr-de Sitter spacetimes. In an idealized
eternal dS space, all other generators should be viewed, as
advocated by Witten, as gauge transformations, which map one
horizon volume into another. Different static observers are
different gauge equivalent descriptions of the same physics.

In a universe which is only asymptotically dS in the future, we
might be interested in these different gauge copies because the
past dynamics of the system might set up different initial
conditions in them.  We might be interested in the fate of
galaxies which we used to be able to see, but which have passed
out of our horizon.  Even in an eternal dS space, a nostalgic
observer might want to learn something about the fate of a friend
who had been foolish enough to let go of her hand, and found
himself swept away by the Hubble flow.

It is extremely important to understand how the Poincare group
emerges from dS space in the limit that the cosmological constant
goes to zero.  The discussion above indicates that it should not
be thought of as the limit of the dS group, most of which consists
of gauge transformations.  Global symmetry generators arise in
General Relativity, by imposing boundary conditions on
hypersurfaces in spacetime.  This is demonstrated quite elegantly
in \gt\ .  The natural hypersurface in dS space is the
cosmological horizon of a given timelike observer.  The
generators which preserve this hypersurface form the $R\times
SO(d-1)$ subgroup of the dS group that we have referred to above.

Near the future cosmological horizon, the metric of dS space takes
the form \eqn\dShormet{ds^2 = R^2 dudv + R^2 d\Omega_{d-2}^2 .}
$v\rightarrow 0$ is the locus of the horizon.The $SO(d-1)$
invariance is manifest, while the static dS Hamiltonian is the
infinitesimal boost of the coordinates $u$ and $v$. This should be
contrasted with the metric of asymptotically flat spacetime near
future null infinity.

\eqn\scriplus{ds^2 = {dudv + d\Omega_{d-2}^2 \over v^2}.} Future
null infinity is the conformal compactification of the
$v\rightarrow 0$ limiting manifold, described by a conformal
structure equal to that of the round sphere plus a single null
coordinate, $u$\asyquant\ . The full asymptotic symmetry group is
the BMS group, consisting of vector fields of the form $f(\Omega
)\partial_u$, semi-direct product with the conformal group of the
sphere (also acting by conformal rescaling of $u$). This large
group arises because relativists want to classify spaces with
classical gravitational radiation in the initial and final states
as asymptotically flat.  In spacetime dimension higher than four,
there is no infrared problem for gravity, and a quantum S-matrix
with finite numbers of particles and no classical radiation,
exists.  In these dimensions one can restrict attention to the
Poincare subgroup of the BMS group (this why we have never seen
the BMS group in perturbative string theory) .   In a conformal
gauge in which the metric on the sphere at infinity is round, the
Poincare subgroup is obtained by restricting $f$ to either $f^0 =
1$ or $f^i = n^i$, the unit vector on the sphere. These
transformations obviously commute, and when account is taken of
the conformal rescaling of $u$, it is easy to see that they
transform as a $d$ vector under the Lorentz group (the conformal
group of the sphere).

It is clear that the only remnant of the dS group, which carries
over to the Poincare group, is the group of $SO(d-2)$ rotations.
The Minkowski translations and boosts arise only in the limit
$R\rightarrow \infty$.  One may be puzzled by the fact that the dS
Hamiltonian is a symmetry generator which is there for all finite
$R$, but seems to disappear in the limit.    We will see below
that in the limit, this Hamiltonian has a degenerate subspace
which becomes infinite dimensional and of infinitely low energy.
The space of states on which the Poincare generators act is
orthogonal to all of these states and matrix elements of all
reasonable measurements (for the asymptotic observer in
asymptotically flat space) between these states and scattering
states, go to zero.   The upshot of this is that the static dS
Hamiltonian does not act on the limiting Hilbert space of
scattering states, which is the space on which the Poincare
generators act.   We will describe the physical basis for this
mathematical behavior below, when we discuss measurement theory in
dS space.

  In asymptotically flat or AdS spaces, one can talk
about idealized measurements on the boundary of spacetime, which
have no effect on the system in the interior, or rather effects
that can be very precisely encoded in the statement that there
are a certain number of incoming and outgoing particles (using
language appropriate to the flat case) of certain types.   The
acts of measuring these particles do not effect what has happened
to them in the interior.   In dS space, no such precise
separation is possible.

The latter statement may seem peculiar to someone who is used to
thinking about the global coordinates for dS space that are
emphasized in \edindia\ and \dscft\ .   Indeed, dS space has a
boundary, past and future null infinity ${\cal I}_{\pm}$ which is
conformal to two spheres.  It is tempting to view data depending
on a finite number of points on each of these spheres as an
analog of the scattering matrix of asymptotically flat, or the
boundary correlation functions of asymptotically AdS, space
times. This would seem to imply an infinite number of states
since, in a semiclassical approximation one can think of an
infinite number of well separated ``particles'' propagating to
the past or the future, with low energies.   Witten has suggested
that this apparent contradiction with the finite number of states
could be resolved if this infinite dimensional S-matrix was not a
unitary operator, but a degenerate matrix of finite rank.   He
also emphasized that these quantities were ``meta-observables"
which could not be measured by any given observer.

I think that there is a much more subtle problem with this
analysis, which has to do with our lack of knowledge of the phase
space of classical gravity.  The phase space of a general
Lagrangian system is the space of solutions of its classical
equations of motion, perhaps restricted by appropriate asymptotic
boundary conditions in the case of field theory.   As noted in
previous sections, it is conventional to describe this in terms of
the field variables and their first time derivatives at fixed
time, invoking the Cauchy-Kowalevska theorem.   Note that such a
description seems to contradict any possible holographic
interpretation of a field theory since we are presented with ``one
degree of freedom per space point''.  This is in fact correct for
non-gravitational theories, but I claim the analysis fails for
theories including gravity.   A first indication of this has been
encountered by numerous people who have thought about the
cosmology of compact universes.   Invoking a Planck scale spatial
cutoff, one would apparently be faced with a change of the number
of degrees of freedom with time, since the dimensions of the
universe expand or contract.

In fact, the Cauchy-Kowalevska analysis is only valid for some
finite time interval and does not discuss the question of global,
nonsingular solutions.  For ordinary field theories the existence
of singularities does not qualitatively change the number of
solutions, but in General Relativity (assuming cosmic censorship)
singularities correspond to the formation of black holes, and thus
to drastic distortions in the geometry of spacetime itself. Given
data on a spacelike slice, it is not easy to specify which
solutions will evolve into black holes.   This problem is
ameliorated if we pose our boundary value problem on the boundary
of asymptotically flat spacetime, and insist that it corresponds
to finite numbers of particles coming in from (going out to)
infinity.  We have a rough idea, in terms of the kinematics of
incoming and outgoing particles, of which configurations lead to
black hole formation.   Quantum mechanically, because of Hawking
radiation, (and ignoring the infrared problem in four dimensions)
even processes that classically form black holes are really
scattering processes involving finite numbers of particles.  So,
in asymptotically flat spacetime, scattering data give us a good
estimate of the number of classical solutions and therefore of the
number of quantum states of the theory.  Note that although this
number is infinite, it is a surface infinity - a holographic
counting of degrees of freedom.

In dS space on the other hand, I claim that the analysis at ${\cal
I}_{\pm}$ is misleading.  If I send in some number of particles
from ${\cal I}_- $ and assume that they do not materially alter
the dS geometry, then I can estimate their energy density in
global coordinates, at the time the dS sphere shrinks to its
minimal size. Obviously, the density becomes larger than Planck
density for some finite number of particles, as long as I do not
make the formal classical approximation of saying that each
particle carries negligible energy, because its classical field is
infinitesimal.   In fact, long before this occurs the geometry
will be distorted.   It is likely that a typical asymptotic
condition on ${\cal I}_-$ leads to a solution with a Big Crunch
singularity.   Similarly, typical data on ${\cal I}_+$ came from
a Big Bang.   If I try to put scattering data on both past and
future, then generically there will be no sensible solution at
all.  More precisely, I believe that the phase space of gravity
coupled to a generic set of physically sensible fields, with the
boundary conditions that the solutions be smooth except for
isolated singularities hidden behind black hole horizons, and the
same asymptotic dS space in the past and future, is
compact\foot{Preliminary results to this effect were obtained in
unpublished work of G.Horowitz and N.Itzhaki.}.

I believe that in this manner, the {\it nonlinear} Einstein
equations are trying to hint to us about the finiteness of the
number of states in AsdS spacetimes\foot{It is often said that
the Bekenstein bound cannot be seen in a classical analysis
because the Planck length goes to zero in the classical limit. In
the above paragraph we have evaded this argument by talking about
classical solutions that carry a finite amount of energy in the
classical limit (what we called particles).  In the standard
classical limit, a particle's Compton wavelength is kept fixed as
$\hbar\rightarrow 0$, so its mass is taken to zero and we could
have an infinite number of particles with finite energy.  A purely
classical statement with the same content, would refer to the
compactness of the phase space with AsdS boundary conditions in
both past and future. } . A rigorous proof that the phase space of
asymptotically dS solutions of Eintein's equations is compact is,
along with Cosmic Censorship for scattering solutions, an
important problem in classical GR whose solution would lend more
credence to the speculations in this paper.

There is another presentation of the semiclassical physics of dS
space, the Euclidean functional integral, that makes the point in
an even more striking manner.  I presented this analysis in
\tbfolly\ but it seems to have been completely ignored. Euclidean
dS space is a sphere.  In the formal\foot{In two space time
dimensions Euclidean functional integral quantization of dS space
is completely rigorous and gives the tree approximation to string
theory.  On the other hand, it has no interpretation in terms of
local physics on the string world sheet.
Polchinski\ref\joeds{J.~Polchinski, {\it The Phase of the Sum
Over Spheres}, Phys. Lett. B219, 252, (1989).} has pointed out
potential problems with the Euclidean formalism in dimensions $>
2$. term in the action.} Euclidean quantization one expands around
the sphere. Rotations of the sphere are diffeomorphisms and one
is instructed to mod out by them since the sphere is compact. The
first correction to the classical saddle point defines free
quantum field theory on the sphere. Analytically continuing this
to Minkowski signature one obtains free quantum field theory on
the static patch of dS space, with metric

\eqn\stat{ds^2 = - dt^2 (1 - r^2 / R^2) + {dr^2 \over (1 - r^2 /
R^2 )} + r^2 d\Omega^2 ,}

\noindent in the thermal state of the static Hamiltonian.   The dS
group, which is the analytic continuation of the rotation group
of the sphere, maps one static patch into another (the static
Hamiltonian and rotation group of the $(d-2)$ sphere leave a
given patch invariant).   The thermal correlation functions can
then be viewed as obtained from a particular dS invariant
Gaussian state of field theory on global dS space, by tracing
over degrees of freedom outside the static patch.

However, the Euclidean path integral prescription instructs us to
think of dS transformations as gauge
transformations\foot{Naively, one is also instructed to treat the
static Hamiltonian and rotation group as gauge transformations.
However Gomberoff and Teitelboim have shown that if we consider
dS space to be the zero mass limit of a dS black hole, then the
Euclidean spacetime has a boundary and these generators are
global symmetries on the boundary.} . {\it Thus the formulation
of the theory in global coordinates contains an infinite number
of gauge copies of the system in a static patch.} The gauge fixed
theory in a single static patch contains all of the physics of dS
space. Note that, although, quantum field theory in the static
patch has an infinite number of states, here the infinite entropy
is associated with large energies, or infinitesimal regions near
the horizon. For energies above the Planck scale, typical
localized states in the static patch are black holes and are not
well described by field theory. Thus, here it is easy to see that
the infinity might be illusory. However, one is led to ask how it
can be that the static patch contains all the physics of dS
space, if the static coordinate system does not cover the entire
manifold?

This question is of course reminiscent of the Black Hole
Information Paradox.  There, a static coordinate system,
appropriate to an observer at infinity also gives rise to a
thermal state.  The region behind the horizon of a black hole is
not gauge equivalent to the rest of the space.  But the resolution
of the Paradox proposed by 't Hooft \ref\tH{C.~R.~Stephens, G.~'t
Hooft, B.~F.~Whiting, {\it Black Hole Evaporation Without
Information Loss },Class.Quant.Grav.11:621-648,1994,
gr-qc/9310006; G.~'t Hooft, {\it Quantum Information and
Information Loss in General Relativity}, gr-qc/9509050; {\it The
Scattering Matrix Approach for the Quantum Black Hole: An
Overview}, gr-qc/9607022, Int.J.Mod.Phys. A11 (1996) 4623. } and
by Susskind and collaborators \ref\stu{L.~Susskind, L.~
Thorlacius, J.~Uglum, {\it The Stretched Horizon and Black Hole
Complementarity}, Phys.Rev.D48:3743-3761,1993, hep-th/9306069. }
has a very similar flavor to our description of dS space.  It
goes under the name of Black Hole Complementarity, and consists
of the claim that the descriptions of physics by external and
infalling observers utilize the same set of states but measure
quantum mechanically complementary observables.

Fischler and I \ref\bfmcosmo{T.~Banks, W.~Fischler, {\it M-Theory
Observables for Cosmological Space-Times}, hep-th/0102077. }
provided a new rationale for this principle and connected it to
the Problem of Time.  We also generalized it to the case of
cosmological horizons. The essential point is that the vector
fields corresponding to time evolution as viewed by these two
observers do not commute with each other. Thus, even
semiclassically, the quantum theories describing the experience
of observers related by a general coordinate transformation, use
non-commuting time evolution operators. It is not surprising then
that physics as viewed by one observer is not quantum
mechanically compatible with physics as viewed by the other.
Indeed, in discussions of the Problem of Time in canonical
approaches to quantizing General Relativity, the idea that there
may be many different Hamiltonians that describe the same physics
is often discussed. The necessity for this point of view is
evident even in the quantization of relativistic particles and
strings, viewed as generally covariant systems in one and two
dimensions.  What is new in low curvature spacetimes with a
horizon is the possibility of having two different {\it
semiclassical} descriptions of the same physical system, which
are quantum mechanically complementary to each other.  Following
't Hooft and Susskind, this would be the correct quantum
mechanical way to describe regions of spacetime which, in the
classical approximation, lose causal contact with each other.
Rather than being described by independent sets of commuting
variables, they are non-commuting descriptions of the same
Hilbert space, each of which separately has a semiclassical
interpretation .

I would claim that the rules of Euclidean Path Integral
quantization of dS space, as adumbrated above, give a
semiclassical derivation of this complementarity (dubbed
Cosmological Complementarity in \bfmcosmo\ )  principle for the
case of a spacetime that is AsdS in both its past and future.
That is, if we imagine a large radius dS space and different
semiclassical observers in it, who are outside each other's
cosmological horizon, the rule tells us that each of them has a
complete description of the physics of all the others.  The
different descriptions are gauge equivalent to each other via the
global dS group.

This is not so interesting in an exact dS space where there is a
symmetry relating local observations of two different static
observers.  Consider however a universe (like our own?) which
began with a Big Bang and asymptotes to dS in the future. There
will again be a multitude of static observers but now they are all
different.  For example, observers in our own galaxy and in some
other galaxy,( say the Sombrero galaxy, ) not gravitationally
bound to us, will, if there is a nonzero cosmological constant,
eventually be outside each other's horizon.  Observers in our
galaxy might well ask where, in their description of the world,
information about the evolution of the Sombrero galaxy is encoded.
There is an obvious answer to this question, suggested by the
earliest studies of black hole physics.

{}From the point of view of a static observer, nothing ever really
goes through the horizon.  That is, the entire region of spacetime
covered by his coordinates ends at the horizon. Instead, as
something approaches the horizon it gets squashed into a smaller
and smaller spacelike region, but never quite disappears. Thus, in
such a coordinate system, it is natural to associate things that
go through the horizon with states localized on the horizon.

It is now important to understand how the number of states
localized on the cosmological horizon of a particular observer
compares to the number that this observer views as localized in
the bulk of spacetime that he can explore.   The key to
understanding this is the Bekenstein-Hawking bound on the entropy
of localized systems.  The largest entropy one can fit into a
region of spacetime is the entropy of a black hole whose horizon
is the boundary surrounding that region.  More properly stated:
when localized energy density is concentrated in a region, if its
entropy is large enough it {\it must} form a black hole with the
requisite radius.

It is well known that the Nariai solution is the maximal black
hole that fits into dS space, and that its entropy is only two
thirds of the dS entropy.   Thus, for a large radius dS space,
the overwhelming majority of quantum states must be viewed by any
given observer as being localized on her cosmological horizon,
rather than on systems localized within the dS bulk. The Nariai
solution represents a very special class of excitations, in which
most of the degrees of freedom of dS space are frozen and the
system explores only a small number of its available states.  By
contrast, the dS vacuum (the thermal Gibbons-Hawking state of the
static observer) and small excitations of it, have a much larger
entropy.  This counting of states seems peculiar to the local
static observer, who in trying to construct a Nariai black hole,
follows her local rule for maximizing entropy.  The resolution of
this tension is the realization that local degrees of freedom are
``stolen" from the horizon.  If we make a small local excitation
we do not substantially alter the bulk of the degrees of freedom,
but the Nariai excitation can only be achieved in a very special
class of states.

For example, if the interpretation of cosmological data in terms
of a cosmological constant is correct then the dS entropy of the
universe is larger by a factor of $10^{25}$ than the entropy of
everything we see as localized excitations (including hypothetical
supermassive black holes in the centers of all large galaxies).
Thus, the classical picture that dS space has an infinitely larger
set of degrees of freedom than what can be seen in a given
horizon, becomes correct in the limit of infinite dS radius, if we
restrict attention to those states that the static observer views
as being localized in the bulk.

The answer to the question : ``What goes on in the Sombrero galaxy
after it has passed through our horizon?" is encoded in the
microscopic quantum state of the cosmological horizon.   This
viewpoint also relieves a certain amount of unease that might be
caused by applying the ideas of Black Hole Complementarity to the
universe. Proponents of complementarity often claim that
observation of the Hawking radiation from a black hole puts the
system in a quantum state which has no classical interpretation
for an infalling observer.  In the black hole context this is
palatable because the infalling observer eventually gets crushed
in the singularity, and there is no compatible notion of
simultaneity for the two observers\foot{In ``nice slice''
coordinates, which try to describe both the interior and exterior
on a surface of simultaneity, the internal observer has to make
measurements with super Planckian time resolution on the slices
where the external observer has absorbed a significant fraction
of the Hawking radiation.  Because of the UV/IR connection she
can no longer be considered a local observer.  This means that
``nice slice'' coordinates do not really exist over time scales
comparable to the time it takes for the external observer to
extract information from the black hole.  The internal part of a
nice slice at these late times, is never well described by local
physics. }. However, we would be disturbed to find that our
measurements could destroy the semiclassical coherence of
observers in the Sombrero galaxy\foot{ Indeed, in dS space, a
perfectly nonsingular \lq\lq Nice Slice \rq\rq, the global
coordinate system, exists. Physics described in this system must
be gauge equivalent to the static patch physics. The global
observer sees nothing happening to the Sombrero galaxy as a
result of our measurements.}. This paradox is resolved because we
do not have the possibility of constructing an apparatus that can
measure such a huge number of states.  If we tried to do so we
would collapse into a black hole. Even if we imagine being able to
measure things by using the microstates of black holes to
construct the measuring apparatus, the maximal dS black hole has
only one third of the entropy of dS space, so for large dS radius
it cannot measure the state of most of the degrees of freedom we
would like to assign to localized systems outside the horizon.

If we restrict attention to well understood measuring devices,
there is a sense in which the global coordinate picture of many
commuting sets of degrees of freedom is valid in the proper
quantum theory of dS space.  Let us use the phrase ``field
theoretic" to refer to states inside each horizon volume which are
well described by quantum field theory. In particular we do not
allow black holes whose size scales to infinity with the dS
radius.  To describe a three dimensional region of size $R$ in
terms of field theory, we must insist on a UV cutoff such that
the typical state (which in field theory means states near the UV
cutoff)has Schwarzchild radius less than $R$.  Thus, in Planck
units,  $M^4 R^3 < R$.  The field theory entropy is of order $M^3
R^3 \sim R^{3/2}$.    In four spacetime dimensions, the total
entropy of dS space indicates the possibility of constructing of
order $R^{1/2}$ mutually commuting sets of ``field theoretic"
degrees of freedom , each of which could describe the field
theoretic states in a given horizon volume. Thus, in the field
theory approximation, {\it i.e.} restricting attention to only
states well described by local field theory (but allowing black
holes with radii much smaller than the dS radius), the picture of
a global dS space is approximately valid unless I try to study
correlations between more that $R^{1/2}$ disjoint horizon volumes
(which I can only do in the far past or future) .

On the other hand, if I construct states with horizon scale black
holes in a single horizon volume then it is no longer possible to
speak of many other independent commuting degrees of freedom. The
horizon size black hole carries a finite fraction of the total
number of degrees of freedom in the system.  The field theory
description of an asymptotically infinite dS space is not far
wrong for large $R$, as long as we restrict attention to low
energy processes.  It is only when a given observer begins to
construct black holes of order his horizon size that he begins to
have a significant effect on the local physics in other horizon
volumes.  If we are not too ambitious, we won't have to worry
about our erstwhile friends in the Sombrero galaxy.

It would be very interesting to put some more mathematical detail
on these arguments by studying multiple black hole solutions in dS
space.  Indeed, even a global coordinate description of single
black hole solutions would illuminate these points.  As far as I
know, no such formulae have appeared in the literature.

In the last few paragraphs, we have touched on the issue of
measurement theory in dS space, which was discussed in
\ref\nightmare{T.~Banks, W.~Fischler, S.~Paban, {\it Recurrent
Nightmares? Measurement Theory in de Sitter Space}, JHEP 0212,
062 (2002), hep-th/0210160.}. Measuring devices in dS space must
be large classical devices, approximately describable by local
field theory, and thus much smaller than the dS horizon size. In
the large $R$ limit, there are two classes of interesting
measuring devices in a given horizon volume: free falling
devices, and devices bound to the measured system at the
origin.   The dS Hamiltonian is the appropriate description of
physics as measured by the latter class of devices.  In the large
$R$ limit, the collection of measurements made by all freely
falling devices, far away from the origin but long before they
fall through the horizon, becomes the Scattering Matrix of the
limiting Minkowski space  (we are talking here of an eternal dS
space with both a past and future cosmological horizon, not an
AsdS space which arose from a Big Bang.).   The approximate
Poincare generators, whose algebra converges to the Poincare
algebra as $R\rightarrow\infty$ are symmetries that act on the
measurements made by the free falling observers.

\subsec{Entropy and the number of states}

In a quantum system, the entropy of a given density matrix is $ -
{\rm Tr} \rho {\rm ln}\rho$.  Up to this point, we have been
conflating the idea of entropy with the logarithm of the number of
states.  This is only valid for the completely uncertain density
matrix on a finite dimensional Hilbert space.  One can have finite
entropy in an infinite system.   The additional semiclassical
input that we need to prove finiteness of the number of states in
dS space , is the fact that the density matrix is thermal, and
that there is an upper bound on the energy spectrum, given by the
mass of the Nariai black hole.   These two facts, combined with
finite entropy, tell us that the number of states is finite, but
not that it is equal to the exponential of the dS entropy.

There are two clues which help us to understand the origin of the
dS temperature.  The entropy of localized excitations of finite
energy is bounded by that of black holes and is, for large $R$,
much smaller than the total dS entropy.  The classical energy of
empty dS space is zero (see {\it e.g.} \gt\ ).  This indicates
that the empty dS entropy should be thought of as coming from
states of zero classical energy, which are the static observer's
view of the world behind her horizon.

I propose that in the quantum theory, the spectrum of the
Hamiltonian at low energies is that of a random $N \times N$
Hermitian matrix $H_V$ with an energy cutoff $\Delta$ of order the
dS temperature.  A possible way of implementing the cutoff might
be to choose $H_V$ from the Gaussian ensemble with covariance
$\Delta$.   The thermal density matrix $e^{- \beta_{dS} H_V}$ is
very close to the normalized projection operator on this $N$
dimensional subspace of low energy states, so that its entropy is
nearly the same as ${\rm ln} N$.  It is in this sense that the dS
entropy is actually a count of the number of states in the quantum
theory.   States with energies higher than the dS temperature have
considerably less entropy, both because they are fewer in number,
and because they are Boltzmann suppressed in the thermal ensemble.
The full Hamiltonian for quantum dS space will have the form $H =
H_V + H_I + H_{loc}$, where $H_{loc}$ gives approximate
eigenstates for localized objects in a single horizon volume and
$H_I$ represents interactions between the vacuum ensemble of
states and the localized states.  One would like to show that
these interactions result in a thermalization of the localized
degrees of freedom, at the dS temperature.   Thus the origin of
the dS temperature will be the dense set of eigenstates whose
dynamics is approximately governed by $H_V$.

To summarize: dS spacetime should be described in quantum theory
by a system with a finite number of quantum states. The positive
cosmological constant is a discrete, tunable parameter, closely
related to the logarithm of the number of states.  Only of order
$R^{3/2}$ of the degrees of freedom can be viewed as local
excitations in a given horizon volume.  This indicates that the
global coordinate picture of many commuting degrees of freedom is
approximately correct for large $R$.  That is, we can rigorously
talk of order $R^{1/2}$ commuting sets of degrees of freedom, each
of which describes field theory in a single horizon volume
(including small black holes).  If the observer in a single
horizon volume tries to make a horizon scale black hole, this
picture is no longer valid and most of the degrees of freedom in
other horizon volumes are also frozen into large black hole
configurations.  For a given observer, the Hamiltonian of dS space
contains a dense and chaotic spectrum of energy levels below the
dS temperature.  These represent the observer's view of things
outside the horizon.   In addition there are localized
excitations, the simplest of which are large black holes.  These
have finite energy, given by the mass parameter in the black hole
solution (for zero angular momentum).   The entropy of dS space is
the thermal entropy of this system, but for the states with no
black holes, it is very closely approximated by the logarithm of
the number of states below the dS temperature.

In fact, the black hole states are not really eigenstates of the
system, except in some approximate sense.  Kerr-dS black holes
evaporate, and most of their decay products fall through the
horizon.  The end product of the decay might be a stable massive
remnant or nothing.  In either case the classical approximation
to the Hamiltonian \gt\ in which there are very high energy stable
eigenstates, must be modified by interactions between the degrees
of freedom associated with the black hole, and those associated
with the horizon, in such a way that the only exact eigenstates
of the Hamiltonian correspond to the dS vacuum and a small number
of stable ``particle" excitations of it.

A more detailed attempt to construct a toy model of dS quantum
mechanics was described in my talks at the Davis Inflation
Conference\ref\davis{T.~Banks, {\it Some Thoughts on the Quantum
Theory of de Sitter Space}, Davis Inflation Conference, U.C.
Davis, April 2003, astro-ph/0305037.}

\subsec{Physics, metaphysics, and mathematics of a quantum dS
universe}

The concept of measurement is essential to all discussions of
physics.  Until very recently physicists thought of their
endeavours as the description of isolated systems. Experimenters
performed measurements on these systems from the outside and
theorists wrote mathematical formulae which were supposed to
predict the results of those measurements.

With the advent of quantum mechanics we have had to pay much more
attention to the concept of measurement, in order to account for
the robust and apparently deterministic nature of measurements in
a world where we believed that all physical systems were subject
to the intrinsically probabilistic laws of the quantum theory.
Starting from the work of Von Neumann\ref\vn{J.~Von Neumann, {\it
The Mathematical Foundations of Quantum Mechanics}, Princeton
Landmarks in Mathematics and Physics Series, 1996.} it has been
argued that the nature of a measurement in quantum mechanics is
correlation of some complete orthonormal set of basis states
$|s>$ of the system with ``pointer states" $|P_s
>$ of the apparatus.  That is, unitary evolution is supposed to
take an initially uncorrelated stated of the combined system,
$\sum a_s |s> |N> $ , into $\sum a_s |s> |P_s >$.  One then tries
to argue that further measurements of system observables in this
correlated state will reproduce their expectation values in the
system density matrix $\rho = \sum |a_s|^2 |s><s|$.   That is,
after the measurement, as long as it remains in interaction with
the measuring apparatus, the system will obey the laws of
classical probability theory with sample space given by the
particular orthonormal basis which has been measured by the
apparatus, and probability distribution given by the square of
its initial wave function.

There has been much discussion of the necessity of including
interactions with a large, random, unmeasured environment to
explain why particular pointer states of the apparatus lead to
decoherence in this manner.   Without quarreling with those
discussions and their applicability to realistic measurements, I
would like to suggest that environmental decoherence is not a
logically necessary component of quantum measurement theory. Large
systems with local interactions ({\it i.e.} quantum field theories
with infrared and (perhaps) ultraviolet cutoffs) provide examples
of systems in which decoherence can occur without a stochastic
environment.  Consider a spin one half particle, and an apparatus
enclosed in a volume $V$ .  The apparatus has two pointer states
which are to be correlated with the $\sigma_3$ eigenstates of the
particle when the two systems come into contact.   We can make a
mathematical model of such an apparatus as a cutoff $\phi^4$ field
theory in the volume $V$, with a potential with two degenerate
minima. We postulate a non-local coupling $\sigma_3 \int \phi (x )
\chi_V( x_p) $ between the two systems. Here $x_P$ is the particle
position and $\chi_V$ is the characteristic function of the volume
$V$.  This non-local interaction is a cartoon of the
amplification system that is required to correlate the state of a
quantum spin with a macroscopic pointer.

Once this correlation has been established, it is very robust.
Further operations that can be modeled by the action of more or
less local operators in the field theory will not detect
interference between the two pieces of the wave function.  In the
limit $V\rightarrow\infty$ we have an exact decomposition of the
Hilbert space into superselection sectors, which never
communicate.  Thus, when $V$ is large in microscopic units, we
can say that an almost classical measurement has been made. Only
tunneling effects, of order $e^{-V}$, are sensitive to the
coherent phases in the correlated wave function.  Thus, a system
is a good measuring device, if the quantum fluctuations of its
pointer observables are analogous to those of a vacuum order
parameter in quantum field theory.  The tunneling time for
fluctuations between different pointer positions is of order the
inverse of the number of states associated with the pointer.

In de Sitter space, there is a bound on the size of a system that
can be described by local field theory.  The tunneling time for
the largest possible field theoretic machine is of order
$e^{R^{3/2}}$, much shorter than the recurrence time in the dS
space\ref\dysonetal{L.~Dyson, J.~Lindsay, L.~Susskind, {\it Is
There Really a dS/CFT Duality}, JHEP,0208:245(2002),
hep-th/0202143; L.~Dyson, M.~Kleban,L.~Susskind, {\it Disturbing
Implications of a Cosmological Constant}, JHEP 0210:011(2002),
hep-th/0208013. }.   Even if we tried to go beyond the well
understood realm of field theoretic machines, and imagined that
we could use localized black hole eigenstates to construct
classical measuring devices we would still find that the
tunneling time between pointer states of such a machine was much
shorter than the recurrence time, because the maximal black hole
entropy is much smaller than the entropy of empty dS space.
Thus\nightmare\ predictions about phenomena on time scales as long
as the recurrence time, have no operational meaning.

More importantly, we learn that many details of the mathematical
quantum theory of de Sitter space, are in principle unobservable.
This means that there will be many mathematical theories that have
the same consequences for all observations, within the bounds of
precision that are allowed by the above arguments.   One should
thus view the quantum theory of dS space as a universality class
of theories, describing the critical limit $\Lambda \rightarrow
0$.  Our considerations of measurement theory suggest that all
members of the universality class should give results for
experiments which do not produce black holes of order the horizon
scale, which are in agreement to all orders in powers of
$\Lambda$.  For such low energy processes, the restrictions on
measurements lead to exponentially small inaccuracies as the
cosmological constant vanishes.

Having established the nature of realistic measurements in a
putative finite dimensional quantum theory of dS space, and the
consequent ambiguity in the mathematical description of this
theory, let us turn to a vexing metaphysical problem raised by
this proposal.   {\it If the number of states is finite, what
determines it?}

There are I think, two possible responses to this question, which
I would call the Anthropic answer, and the Pythagorean answer.
Actually, the Pythagorean answer will be seen to require a very
weak form of the anthropic principle as well. The Anthropic answer
invokes the results of Weinberg \ref\wein{S.~Weinberg, {\it
Anthropic Bound on the Cosmological Constant}, Phys. Rev. Lett.
59, 2607, (1987).} to claim that a value of the (positive)
cosmological constant larger than what is observed would lead to
a universe devoid of galaxies and thus (presumably) of living
organisms of any kind. Actually the galaxy bound exceeds the
observed value by a factor of order $100$ and one must resort to
arguments about what a ``typical" universe obeying the bound
would look like. Intrinsic to any such discussion is an {\it a
priori} notion of what the ensemble of possible values of
$\Lambda$ is and what the probability density on this ensemble
looks like.  The arguments of Weinberg assume a fairly flat
density in the vicinity of $\Lambda = 0$.  If on the other hand,
we associate $\Lambda$ with the number of states, then the small
$\Lambda$ region is the region with a large number of states and
a flat probability density near vanishing cosmological constant
is assuming a cutoff on the number of states.   This does not
appear reasonable, and puts in a scale by hand.

Cosmological SUSY breaking can help to solve this problem.  It
implies that systems with a large number of states become more and
more SUSic.  In the limit of small gravitino mass, atoms and
nuclei can decay to a bose condensed ground state by gravitino and
photino emission.  Life is impossible in a very SUSic universe.
Note that, although this argument uses life of our type as its
basis, it may still be a weak anthropic argument.  It is possible
that the constraints\ref\susypheno{T.~Banks, {\it The
Phenomenology of Cosmological SUSY Breaking}, hep-th/0203066.} on
the limiting SUSic vacuum of dS space are so strong that there is
only one solution and it predicts the low energy supersymmetric
standard model with all of its parameters. Thus, purely
mathematical arguments might lead to the unique conclusion that
nuclear physics and chemistry are as they are in the real world,
whenever the cosmological constant is small. This would determine
the possible types of life. The precise value of the cosmological
constant would then be predictable only by anthropic arguments.

Given the latter assumption we can get even stronger lower bounds
on $\Lambda$ by combining CSB, the anthropic principle, and the
assumption that the weak scale is determined by the scale of SUSY
breaking . If $\Lambda$ is too small then the weak scale will be
so small that the dominant contribution to the proton neutron mass
difference will be electromagnetic\foot{This argument about the
effect of lowering the weak scale is due to Dimopoulos and
Thomas.}. Protons will decay rapidly due to the relatively strong
weak interactions, and there will be no atoms or heavy nuclei.
Dimopoulos and Thomas \ref\dt{S.~Thomas, Private Communication}
estimate that the weak scale can be no lower than a factor of
three smaller than its actual value, to prevent this disaster.
According to CSB, the weak scale vanishes like $\Lambda^{1/8}$,
so we have an anthropic lower bound on $\Lambda$ which is about a
factor $10^{-5}$ smaller than its ``real" value.  Because of the
weak power law dependence of $M_W$ on $\Lambda$ it seems unlikely
that refinements of these arguments could produce a really tight
lower bound on $\Lambda$.

A possible way to do better would be to take up the suggestion I
made in \susypheno\ that some of the small parameters in the quark
mass matrix could be functions of $\Lambda$ as well.  The
motivation for this is that the discrete $R$ symmetry which
guarantees Poincare invariance of the limiting SUSic theory, and
which is broken by interactions with the horizon states in dS
space, might be related to the discrete flavor symmetries which
constrain the quark masses.   If this were the case for the up
and down quark masses, then the proton neutron mass difference
would be a more rapidly varying function of $\Lambda$.  If we
assume the entire ratio of the up to top quark masses is due to a
power of $\Lambda$, then the proton neutron mass difference
scales like $\Lambda^{1/6}$ and the anthropic lower bound on
$\Lambda$ is about $10^{-3}$ smaller than its real value.

As an aside, I should mention that although the anthropic
determination of $\Lambda$ is often considered a great success
for the anthropic principle, there are a large number of hidden
assumptions in such a statement.  We have already seen that the
apparently innocuous assumption of a flat probability distribution
near $\Lambda = 0$ does not make much sense if we think of
$\Lambda$ as a parameter controlling the number of states in the
quantum theory.  We had to use the additional assumption that any
$\Lambda =0$ theory was exactly SUSic to make any use of the
anthropic principle with the more plausible assumption of a
uniform probability distribution on the number of states.   Even
then we had to use the hypothesis of CSB and even more
speculative assumptions about dependence of low energy parameters
on $\Lambda$ to get a reasonably tight lower bound.  Furthermore,
Weinberg's upper bound on $\Lambda$ assumes that the dark matter
density at the beginning of inflation, and the amplitude of
primordial density fluctuations are fixed to there values in the
real world.  In most attempts to motivate anthropic arguments by
assuming an elaborate potential energy landscape and the ability
to jump between local minima (a picture that is on very shaky
grounds if one believes the arguments of this paper), both of
these parameters would be expected to fluctuate randomly.  The
anthropic prediction for the central values in this
multidimensional parameter space is not very impressive.   Thus,
one really needs a theory like CSB where {\it only} the
cosmological constant is allowed to vary, to claim that the
anthropic determination of $\Lambda$ is successful.

Perhaps the most attractive feature of the anthropic argument is
that it does not require us to know much about the Meta-theory,
which determines the probability distribution of the cosmological
constant.  One requires only that such a theory exists and that
the probability distribution in the vicinity of the anthropic
bound is nonzero, and reasonably smooth.  The lack of dependence
on details of the Meta-theory is important, because it is unlikely
that any of those details could be checked by experiment.  If we
needed to understand an elaborate mathematical theory, most of
whose structure could never be tested, in order to believe in the
anthropic bound, then that bound would appear much less plausible.

The Pythagorean answer to the question of the number of states,
is an attempt to build a Meta-theory using number theoretic
concepts.   One imagines a ``universe machine'', which (in a time
which has nothing to do with any time coordinate in our universe)
spits out some number $n$ of commuting Pauli spin operators (or
some other elementary quantum system like the spinor variables of
the holographic cosmology described below) and allows them to
interact according to the rules of quantum asymptotically de
Sitter space-time.   $n$ is to be chosen by some number theoretic
criterion, built into the (hypothetically elegant) structure of
the universe machine. An ideal Pythagorean solution would find
$n$ to be uniquely defined by some simple criterion.  For
example, if Fermat's theorem were false, and had a unique but
huge counterexample, it might have fit the bill. What is required
is a number theoretic problem that has a unique solution, which
happened to be the value of $n$ that fits the cosmological
constant.  If such a problem could be found, we might believe in
the theory even if no other experimental checks of the mechanism
behind the universe machine could be done.

A more plausible construction might rely on a number theoretic
problem that had sparse solutions.  As an example, we might
consider the requirement that $n$ be a Mersenne prime.  These are
primes of the form $2^k - 1$, and there are only $39$ of them
known.  The largest has $k$ of order $13.5 \times 10^6$.  The
resulting value of $n$ is much too large to fit the observed value
of the cosmological constant.  Indeed, there are no Mersenne
primes which fit the right value, all giving answers which are
much too large or much too small\foot{The two closest Mersenne
primes predict a cosmological constant of order $10^{-157}$ or
$10^{-38}$.}. However, one can imagine a similar problem in number
theory, which, for some value of $k$ hit the cosmological constant
on the nose, and missed by a huge margin for all other values of
$k$. We could then used this elegant number theoretic machinery to
construct a much more satisfying version of the anthropic
argument.  The {\it a priori} probability distribution for the
cosmological constant would have point support and all but one of
the points would violate the anthropic upper and lower bounds that
we have described, by large amounts. Thus, a Pythagorean choice
of $N$ would ultimately depend on anthropic arguments, but would
be much more compelling than the argument of Weinberg.  There
would be a very sparse set of choices for this fundamental
integer and only one of them would be compatible with a very weak
version of the anthropic principle.

I must admit to a great deal of unease in talking about these
arguments.   Consider the following model of a Meta-theory: A
supreme being plays dice with himself, and on the basis of each
throw, decides to construct a universe with a finite number of
quantum states obeying the famous, yet to be constructed, rules
for quantum cosmology in such a universe.   Only the number of
spins $n$ is decided by the throw of the dice.  We then apply the
anthropic argument.   As theoretical physicists, we would
certainly find an elegant mathematical model of a Meta-theory
more satisfying than the supreme being model, but our inability to
perform experiments for the values of $n$ that are ruled out by
the anthropic argument, leaves us with no experimental proof that
the supreme being model is any less right than the mathematical
one.   We must ask ourselves whether we are really doing science.
So must anyone who indulges in anthropic speculation.

\newsec{\bf Supersymmetry}

\subsec{Breaking SUSY on the horizon}

It is clear that dS space violates SUSY.  There is a dS analytic
continuation of the AdS SUSY algebra, but it has no unitary
representations and is not compatible with quantum mechanics. The
basic problem is that the dS group has no highest weight
generators (it is isomorphic to the Lorentz group in an
appropriate number of dimensions) and so no bosonic generator can
be written as a positive product of supercharges.   Furthermore,
if we believe the arguments above, the dS group itself should be
viewed as a group of gauge transformations, with the coset of the
static subgroup not acting on the Hilbert space of states of a
given static observer. The question is, by how much does dS space
with a given cosmological constant violate SUSY ?

This question touches on a nastier one, namely what are the
precisely defined mathematical observables in dS space, or are
there any at all?  I have discussed this briefly in a previous
section, but it is beside the main point.   Whatever the precise
definition of observables in AsdS spaces, it must be true that
there is some approximate notion of low energy physics described
by an effective Lagrangian.  In this context, breaking of SUSY can
always be described as spontaneous, as long as the gravitino mass
is much smaller than the Planck scale.  The hypothesis of
CSB\tbfolly\ guarantees that this is so, for it links the SUSY
breaking scale to a positive power of the cosmological constant,
which, according to the hypothesis, is a tunable parameter.
Furthermore, since the limiting theory, with vanishing
cosmological constant, is supersymmetric, the goldstino must be
part of a linear supermultiplet, and SUSY breaking must be
described by some standard (or novel) low energy mechanism. The
novelty of the current approach is that one is led to accept the
existence of what appear to be fine tuned relevant parameters in
the low energy effective Lagrangian.   For someone like myself,
who has spent a good part of his career in physics searching for
dynamical explanations of mass hierarchies, this seems like a
revolting and reactionary approach to the problem.

The crucial point however is that the dynamics which explains
these finely tuned numbers is, according to CSB, a {\it new
critical dynamics} of the large set of degrees of freedom that
become available as the cosmological constant goes to zero. In
thinking about quantum gravity, we have grown used to invoking
various kinds of infrared critical behavior (asymptotic freedom
and nontrivial fixed points).   We have imagined that since the
theory contains an apparent UV cutoff scale, the Planck mass, that
there was no problem of an ultraviolet infinity of degrees of
freedom.  I have stated above, that this is wrong. In fact, the
UV/IR connection, as manifested in the black hole spectrum in
asymptotically flat space suggests that the theory with vanishing
cosmological constant indeed has a UV critical problem, of a type
never encountered before.  This is a critical problem to which the
standard paradigm of Lorentz invariant conformal field theory (in
space time) simply does not apply. In \tbfolly\ I invoked this new
critical phenomenon as an explanation for the change of the
exponent, $\alpha$, in the relation, $m_{3/2} \sim
\Lambda^{\alpha}$ (Planck units), between the gravitino mass and
cosmological constant.  This led me to suggest virtual black holes
as the mechanism which renormalizes the critical exponent.

I no longer believer this argument.  Although it is easy to argue
for unsuppressed virtual black hole production in the high energy
parts of Feynman graphs, the probability of tying all the black
hole decay products back into a single particle (in order to
renormalize the gravitino mass) seems very small.  More
importantly, I realized that, as discussed above, black holes
seen by a static observer account for only a tiny fraction of the
states of dS space.  Indeed, there is another, peculiar, IR
critical phenomenon that occurs as the cosmological constant goes
to zero.  In the holographic picture of dS space, the states on
the cosmological horizon must be not only degenerate, but have
very small eigenvalues of the static Hamiltonian.  This is
required, in order for them to participate in thermodynamics at
the (very low) Hawking temperature.  In my current view, it is
these states that are responsible for the relatively large
renormalization of the gravitino mass.

The classical low energy Lagrangian for SUGRA coupled to chiral
and abelian vector superfields has a potential of the form

\eqn\pot{V = e^K [F_i \bar{F}_{\bar{j}} K^{i \bar{j}} - 3 |W|^2]
+ D_a^2 , } where $K$ is the Kahler potential and the $F$ and $D$
terms have their usual expressions in terms of the chiral fields.
If the cosmological constant is tuned to zero, the gravitino mass
(interpreted as the mass of a scattering state in asymptotically
flat space) is given by $\sqrt{e^K |F|^2}$ , where we have
introduced a shorthand for the norm squared of the one form $F_i
dz^i$ in the Kahler metric.  If we turn on a positive cosmological
constant this term no longer has the same meaning, but it still
serves as the coefficient of the nonderivative quadratic term in
the gravitino Lagrangian.  We will take it as our estimate for the
size of SUSY breaking.  It is clear that this Lagrangian has no
{\it a priori} connection between the size of SUSY breaking and
the cosmological constant.  However, a cosmological constant ,
$\Lambda$ much smaller than the SUSY breaking scale is achieved
only by subtracting two terms, each of which is many orders of
magnitude larger than $\Lambda$.  Quantum corrections in field
theory seem to restore the problem solved by the classical fine
tuning.

A number of authors\ref\hcknt{A.~Cohen, D.~Kaplan, A.~Nelson,
{\it Effective Field Theory, Black Holes, and the Cosmological
Constant}, Phys. Rev. Lett. 82, 4971, (1999), hep-th/9803132;
S.~Thomas, {\it Holography Stabilizes the Vacuum Energy}, Phys.
Rev. Lett. 89, 081301, (2002).} have suggested that these
calculations are faulty because many of the intermediate states
used in these calculations suffer large gravitational
distortions. In particular, Cohen, Kaplan, and Nelson propose a
calculational procedure in which the radiative corrections to the
cosmological constant are consistent with observational bounds,
and normal particle physics calculations are affected at a level
below, but in some cases close to, current experimental
precision. Thomas has proposed a different way of modifying the
results of QFT calculations.  In contrast to CKN he uses only the
holographic bound on states, and does not predict such large
corrections to other low energy calculations.

These calculations show how plausible holographic constraints on
the field theory formalism can remove the technical problem of
loop corrections to the fine tuning, but the problem is deeper
than that.  The effective Lagrangian formalism is more background
independent than we have argued the fundamental quantum theory of
gravity has any right to be.  It treats asymptotically flat, dS
and AdS spaces as part of the same theory and the cosmological
constant as a calculable parameter.  Quantum mechanically, these
systems have very different structure: dS space has a finite
number of states, while the infinite AF and AdS Hilbert spaces
have a radically different behavior of the high energy density of
states.

The cosmological constant is always a regulator of the growth of
the number of states at high energy.  It is a strict UV cutoff in
dS space and a crossover scale between superexponential and
subexponential growth of the density of states in AdS space (with
the AF case viewed as the limit where this scale goes to
infinity).  As such it seems like a fundamental parameter of the
theory rather than a parameter which can suffer renormalization.
Indeed, as I have emphasized, in AdS/CFT the cosmological constant
in Planck units is identified with a power of $N$, a fundamental
integer characterizing the quantum theory, rather than an
effective parameter.

{}From this point of view, the fine tuning of the cosmological
constant in the effective theory seems to be merely a way of
putting this fundamental piece of high energy information into the
theory.  It is not that different than imposing a symmetry whose
justification comes from the high energy theory. What about the
relation between the cosmological constant and the SUSY breaking
scale?

Our discussion of the structure of Hilbert space in a dS spacetime
leads us to consider contributions to the renormalization of the
gravitino mass coming from diagrams like those of Fig. 1, in which
the internal gravitino line propagates out to the cosmological
horizon, and back.  The external gravitino lines are localized in
a small, approximately flat region of spacetime whose scale is
somewhat bigger than the gravitino Compton wavelength.

In propagating out to the horizon, the gravitino line enters what,
for the static Hamiltonian, is a region of infinitely high
temperature. It is able to interact with the mysterious horizon
states.   While amplitudes like these are suppressed by $e^{ - c
m_{3/2} R}$ due to virtual propagation over a large spacelike
distance, there is a potential contribution from interaction with
of order $e^{R^2 M_P^2}$ near horizon states.  Until we understand
how to compute these interactions, we cannot claim that these
renormalization effects are small, or estimate how they depend on
$R$.

In \susypheno\ ,I argued that the only sort of conventional low
energy SUSY breaking mechanism consistent with CSB and some
rudimentary standard model phenomenology was one in which a
combination of F and D term constraints for fields charged under
a new $U(1)$ gauge theory with Fayet-Iliopoulos (FI) term , leads
to a vacuum with spontaneously broken SUSY. It remains to be seen
whether a phenomenologically viable model of this type can be
found.  In the meantime I have discovered another class of
models, based on dynamical SUSY breaking, that also seems to be
consistent with CSB.

Since my understanding of the phenomenological implications of CSB
is much less complete than I had originally thought, I will not
review this work here.  However, there are a few points that are
essential to the following argument.  Despite the fact that dS
space breaks SUSY, at the level of the low energy effective
Lagrangian this breaking must appear spontaneous, unless the scale
of the gravitino mass is above the Planck scale.  This follows
from the necessity of SUSY Ward identities to any low energy
theory of gravitinos interacting with gravity.  Since SUSY is
local, we can always make its breaking look spontaneous, by
introducing a nonlinear Goldstino field.  Moreover, by assumption
we have a one parameter set of theories with a SUSic limit.  Near
the limit, the scale of SUSY breaking goes to zero, and the
Goldstino must actually fit into a linear SUSY multiplet. Finally,
to naturally assure a zero cosmological constant limit, we invoke
a discrete complex R symmetry of the limiting theory.

The discrete R symmetry is {\it explicitly} broken by terms that
vanish with $\Lambda$.  Thus, we are led to postulate an R
symmetric low energy Lagrangian with a Poincare invariant, SUSic,
vacuum state. When explicit R breaking terms (including a constant
in the superpotential that we can use to fine tune $\Lambda$) are
added, the model must spontaneously break SUSY.

Our goal is to estimate the size of the R breaking terms that are
induced by interactions with the cosmological horizon.  These are
given by Feynman diagrams with vertices localized near a given
static observer, and lines carrying R-charge out to the horizon.
The gravitino always carries R-charge, and in most models will be
the lightest R charged particle.  We will see that graphs with
gravitino lines going to the horizon are the important ones.
Consider a graph with one such line (Fig. 1).  As noted above, it
will contain a factor $e^{- m_{3/2}R}$ , where $R$ is the
spacelike distance to the horizon.

There is a set of arguments that recovers the relation $m_{3/2} =
M_P (\Lambda /M_P^4)^{1/4}$ from such configurations.  First we
must assume, based on the area scaling of horizon entropy, that we
can view the states on the horizon as distributed uniformly over
it.   Then the interaction of some localized particle with the
horizon, would be only with those states concentrated in the area
of the horizon explored by the particle.   A particle of mass $m$
can move a proper distance of order $1/m$ along a null surface
like a horizon.  For longer proper distances it must be
considered to follow a timelike trajectory and cannot stay in
contact with the horizon. However, from the point of view of a
particle which must return to the observer at the origin (as must
be the case for the virtual line in Fig. 1) the horizon is a very
hot place, and the particle undergoes strong interactions as it
moves around near the horizon.  In particular, it's position is
subjected to a random kick every time it moves a Planck
distance.   So its motion should be viewed as a random walk, and
in a proper distance $1/m$ it moves only a distance $\sqrt{1/m}$
from its starting point. Thus, it explores an area $\sim
1/m$\foot{These estimates are done in four dimensions.  The
general exponent is $A \sim m^{1-d/2}$.}. Note that these random
kicks only move the particle within the horizon.  In the static
frame there is a huge inertial potential energy which pins it to
the horizon.  And if it needs any repetition:  in the holographic
description there is no place outside the horizon.

To understand the order of magnitude of the correction we have to
say something about a model of physics at the horizon.  The area
scaling of entropy might suggest a cutoff field theory model, a
sort of quantization of the fields of the membrane paradigm. This
I believe to be wrong.  Consider for example a cutoff field
theory model of a black hole horizon.  Along with an area's worth
of entropy, it would predict an area's worth of energy density.
But a black hole's area scales with a power of the energy greater
than one, so this is inconsistent.  In a field theory, the
splittings between horizon states would scale like an inverse
power of the area.  Instead, we expect the splittings between
levels of a black hole to be exponentially small in the area (the
inverse of the spectral density) rather than the power laws that
would be predicted by a field theory.

A better model can be constructed in 4 spacetime dimensions
(which may be the only place we need it for dS space). The horizon
is a two sphere.  Consider free nonrelativistic fermions
propagating on this two sphere, in the presence of the background
magnetic field of a monopole at the center of the sphere.  The
fermions are doublets of an $SU(2)$ isospin symmetry and their
isospin operators do not appear in the Hamiltonian.  Take the
monopole charge to be large, and consider a completely filled
lowest Landau level.  The isospin degeneracy gives a number of
degenerate states exponential in the area of the sphere, in units
of the quantized Larmor area, which should evidently be of order
the Planck area. This will be our model of the degenerate horizon
states.

It is well known that there are linear combinations of single
particle states in the first Landau level which are approximately
localized within a quantized Larmor area. Particles in the bulk of
dS space will be assumed to interact with these horizon states
via a localized function of the difference between the particle
coordinate on the sphere and the fermion guiding center
coordinates.   The localization length of this function is of
order the Planck length.  The interaction is again independent of
the fermion isospin operators.

With this model of horizon states and our description of the
random walk of bulk gravitinos on the sphere, it is clear that the
gravitino interacts with of order $e^{1/m}$ states. In this simple
model, the interaction is insensitive to the degeneracy and so we
sum coherently over the degenerate states. More generally, if the
contributions of a finite fraction\foot{Finite means finite as
$R\rightarrow\infty$. Actually, we want the fraction to vanish as
a power of $R$, but this is a subleading correction to the
exponential terms we are studying.} of these states to the
renormalization of the effective Lagrangian adds coherently, then
we will have

\eqn\chglag{ \delta {\cal L} \sim e^{-m R} e^{1/m} }

Only if $ m \sim R^{-1/2} \sim \Lambda^{1/4}$ will this
contribution be neither exponentially growing or vanishing as
$R\rightarrow\infty$.

I would like to view this as part of a self consistent
calculation of the gravitino mass, in the spirit of the
Nambu-Jona-Lasinio model.  That is, we add R violating terms to
the low energy Lagrangian which induce a gravitino mass because
they lead to low energy spontaneous SUSY breaking. Then we ask
for what value of this mass the R violating terms will be induced
by diagrams like that of Fig. 1.  If our crude estimates of the
exponential enhancement are correct, then only the scaling
exponent $1/4$ is self consistent. If the gravitino mass were
smaller than this, then our estimate would give an exponentially
{\it larger} contribution, so a smaller mass is not consistent.
Similarly, a larger mass, would predict an exponentially small
contribution from the horizon.  However, we know that there are
no other sources for large $\Lambda$-dependent renormalizations
of $m_{3/2}$. So a larger mass is also inconsistent.

Why should we imagine that the above estimate applies to the
gravitino mass, rather than that of some other state?  At the
moment, my only argument is based on the low energy effective
Lagrangian.  In \susypheno\  I argued that this had to be a system
which was an R violating relevant perturbation of a
supersymmetric, R symmetric system. In that context, there are no
SUSY splittings that are smaller than the gravitino mass\foot{In
fact in the explicit models constructed in the above reference,
all other SUSY splittings are much larger than the gravitino
mass.}. Furthermore, it is clear that the contribution from the
horizon is dominated by interactions with the lightest particle
available. The particle in Fig. 1 must carry R charge, in order
to generate R violating interactions. It is very plausible
phenomenologically, and certainly true in the models of
\susypheno\ that the gravitino is the lightest R charged particle.

Together, these arguments suggest a radical change in our thinking
about the cosmological constant problem.  That problem itself is
insoluble (see however the section on Metaphysics above) - the
cosmological constant in Planck units is simply the inverse
entropy of the universe, and is built in to the structure of the
Hilbert space.  The crucial relation between the scale of SUSY
breaking and $\Lambda$ depends on physics that we do not have good
control over, but I have argued that the eventual quantum theory
of dS space will give a value for the SUSY breaking scale which is
compatible with observations.

\subsec{Uniqueness of the limiting SUSic vacuum}

A crucial question for the program described in this paper is the
uniqueness of the limiting SUSic vacuum state.  Our claim is that
there is a quantum theory of dS spacetime with any finite number
of states and that the zero c.c. limit of these theories must be
a Super Poincare invariant, R invariant system (with an invariant
vacuum state) satisfying several other properties.  Nothing in
our argument tells us that there cannot be many such systems,
perhaps even an infinite number.

This would be something of a disaster for our program.  It is
likely that the low energy dynamics of any theory satisfying our
criteria would be sufficiently complicated that we would have
little chance of deciding whether complex, intelligent organisms
could evolve in these alternative universes.  There would be many
theories of quantum gravity in asymptotically dS space with a
fixed value of the cosmological constant, and so our theoretical
framework would not be terribly predictive.  Many key features of
the world we know, like the choice of the low energy gauge group,
would be random accidents, correlated to the structure of our
particular form of life, but having no explanation at a more
fundamental level.  The best we could hope to do would be to find
a theory in this class with the right low energy gauge group, and
calculate the parameters in the standard model from first
principles. The task of theoretical physics would end with the
establishment of the existence of a theory which contained the
right gauge groups and the right parameters.

It is to be hoped then that the limiting SUSY theory is unique or
is member of a small finite family.  Infinite sets of theories
with low energy dynamics trivial enough to rule out the
possibility of complex organisms would also be acceptable.

The most important question is to find an algorithm, which will
generate all possible theories with isolated SUSic vacua. Here I
want to suggest something which seems to run contrary to (but is
in fact completely compatible with) my claim that off shell
effective potentials are not sensible objects in string theory.
If we look at compactifications to four dimensions with four
supercharges, there are a variety of regions in which we seem to
obtain moduli spaces of limiting theories with exact
super-Poincare invariance.   Examples are the heterotic string,
or large radius compactification of 11D SUGRA on $G2$ manifolds
or $CY_3 \times S^1/Z_2$.  For finite values of the string
coupling or radial moduli we expect a superpotential to be
generated.   Furthermore, the superpotential has an expansion in
$e^{-aS}$ for some $a$, which can, in principle, be computed by
instanton analysis. Moreover, there is every indication that
these expansions have a finite radius on convergence. It would
seem that these expansions unambiguously define a section of a
locally holomorphic line bundle on some kind of complex moduli
space.

We cannot say the same about the Kahler potential, which is given
by a divergent series.   However, the question of the existence of
Super-Poincare invariant vacuum states is independent of the
Kahler potential, and indeed has the same answer for any choice
of the connection on the line bundle.  One is led to conjecture
the existence of a topological version of string/M theory with
$4$ SUSYs, which would define this complex moduli space in a
non-perturbative fashion, and compute the number of solutions to
the equations $W = dW = 0$.  Perhaps it will also give answers to
questions about the spectrum of massless states at each solution.
Techniques for answering the analogous question about the
spectrum of BPS branes and their world volumes superpotentials in
compactifications to four dimensions with eight supercharges have
been developed by Douglas and others \ref\dougoth{E.~Witten, {\it
Chern-Simons Gauge Theory as a String Theory}, hep-th/9207094;
M.~R.~Douglas, B.~R.~Greene and D.~R.~Morrison, {\it Orbifold
resolution by D-branes}, Nucl.\ Phys.\ B506, 84 (1997),
hep-th/9704151; I.~Brunner, M.~R.~Douglas, A.~E.~Lawrence and
C.~Romelsberger, {\it D-branes on the quintic} JHEP 0008, 015
(2000),hep-th/9906200; S. Kachru, S. Katz, A. Lawrence, J.
McGreevy, {\it Mirror symmetry for open strings}, hep-th/0006047;
F.~Cachazo, B.~Fiol, K.~A.~Intriligator, S.~Katz and C.~Vafa,
{\it A geometric unification of dualities}, Nucl.\ Phys.\ B628, 3
(2002),hep-th/0110028; M.~R.~Douglas, S.~Govindarajan,
T.~Jayaraman and A.~Tomasiello, {\it D-branes on Calabi-Yau
manifolds and superpotentials}, hep-th/0203173.}.

One would hope to be able to do more, that is to find a
constructive prescription for computing the full scattering matrix
(here I abuse language and neglect the infrared problem ) at each
of these solutions. Thus, the goal that I propose is to find an
algorithm for constructing all Minimally Super Poincare invariant
S-matrices for quantum gravity in $4$ dimensions \foot{Again, I
am using S-matrix as a shorthand for a more sophisticated object,
which includes soft graviton bremstrahlung.} . They are rare
jewels and constructing a machine that produces them would seem
to be a worthy goal for much future research in string theory.  I
believe that it is of great interest even if one rejects the
wilder conjectures in this paper.

\subsec{SUSY and the Holographic Screens: An Idea for An Idea}

The current section is even more speculative than the rest of this
paper. Given the importance of supersymmetry in string theory,
one is led to ask whether there is a more geometrical way of
understanding the necessity for incorporating supersymmetry into
a quantum theory of gravity.  In my lecture at the Strings at the
Millenium Conference in January of 2000\ref\tbmill{T.~Banks, {\it
Supersymmetry and Space Time} Talk given at the Strings at the
Millenium Conference, CalTech-USC, January 5, 2000.} , I
suggested that the geometrical origin of local SUSY was the
holographic principle.  The formulation of the holographic
principle for asymptotically flat
spacetimes\ref\thornsuss{C.~Thorn, {\it Reformulating String
Theory With The 1/N Expansion}, hep-th/9405069; L.~Susskind, {\it
The World as a Hologram}, J.Math.Phys. 36 (1995) 6377,
hep-th/9409089.} makes it apparent that the choice of a
holographic screen (holoscreen) is a gauge choice. In extant
formulations in asymptotically flat space-times, the holoscreen
is chosen to be a light plane, corresponding to a particular
choice of light front gauge.

Bousso's\ref\raph{R.~Bousso, {\it A Covariant Entropy Conjecture
}, JHEP 9907 (1999) 004, hep-th/9905177 ;{\it Holography in
General Space Times}, JHEP 9906 (1999) 028, hep-th/9906022; {\it
The Holographic Principle for General
Backgrounds},Class.Quant.Grav. 17 (2000) 997, hep-th/9911002.}
general formulation of the holographic principle makes it
abundantly clear that there are many ways to project the data in
a given spacetime into collections of holoscreens.  This suggests
that any quantum theory of gravity based on the holographic
principle must have a new gauge invariance, going beyond (but
intertwined with) general coordinate invariance.  It was obvious
that this must be related in some way to local SUSY.

The mathematical connection comes from the Cartan-Penrose (CP)
equation, relating a null direction to a pure spinor.  In fact, a
pure spinor not only defines a null direction, but also a
holoscreen transverse to that direction.  One can think of the
choice of a pure spinor at each point in space-time as a choice
of a holographic screen on which the data at that point is
projected.  A local change of the choice of spinor corresponds to
the holoscreen gauge invariance noted above, and smells like it
has something to do with local SUSY.

The spinors in the classical CP equation are bosonic, but
projective.  Nothing in the correspondence depends on the overall
complex scale of the spinor.   The current,
$\bar{\psi}\gamma^{\mu}\psi$ constructed from the pure spinor, is
a null direction, again a projective object.  The equation knows
only about the conformal, and not the metrical structure of
space-time.  In particular, although the classical CP equation
makes a natural connection between the choice of a pure spinor at
a point in spacetime, a null direction, and the {\it orientation}
of a holographic screen, it says nothing about how far the screen
is from the point, nor the geometrical extent of the screen.  The
geometrical interpretation of pure spinors originated in the work
of Cartan\ref\cart{E.~Cartan, {\it The Theory of Spinors}, MIT
Press, Cambridge MA, reprinted from the original French edition,
Hermann, Paris, 1966.}.  In four dimensions, Penrose described
the geometry with the picturesque phrase ``A spinor is a flagpole
(null direction) plus a flag (holoscreen)."\ref\mtw{C.W.~Misner,
K.S.~Thorne, J.A.~Wheeler, {\it Gravitation}, 41.9, p. 1157,
W.H.Freeman, San Francisco, 1973, and references therein.}.

Below we will turn the CP spinors into quantum operators.  The
quantization procedure breaks the classical projective invariance
of the CP equation, leaving over a discrete phase invariance.
This should be viewed as an additional gauge symmetry of the
quantum theory.  We will show that, after performing a Klein
transformation using a $Z_2$ subgroup of this gauge symmetry, the
CP spinors become Fermions and $Z_2$ is just $(-1)^F$.

The motivation for this particular method of quantizing the
classical variables, combines the classical relation between
spinors and holoscreens with the BHFSB relation between area and
entropy.   In this way the breaking of projective invariance is
precisely equivalent to the introduction of metrical structure on
space-time.  The detailed answer to the question of the size and
location of the holoscreen is thus quantum mechanical.  As we
will see, it also depends on {\it a priori} choices of boundary
conditions for the space-time and on a choice of gauge.  Below,
we will concentrate on non-compact, 11 dimensional cosmological
space-times which expand eternally.

\def\hn{{\cal H}_n}

\subsec{Quantizing the CP equation}

As explained above, the connection between local SUSY and
holography comes via the Cartan-Penrose relation between spinors
and null directions. Given a null direction $p^{\mu}$ in $d$
dimensions, there are always $2^{[d/2] - 1}$ solutions of the
Cartan-Penrose equation

$$p_{\mu} \gamma^{\mu} \psi = 0$$.

Furthermore, solutions of this equation for general $p^{\mu}$ can
be characterized as the submanifold of all (projective) spinors
satisfying $\bar{\psi} \gamma^{\mu} \psi \gamma_{\mu}\psi = 0$.
The null direction, of $p^{\mu}$ is determined by this equation,
but since the spinor is projective, not its overall scale.

 A choice of pure spinor completely determines a
null direction and a $(d-2)$ dimensional hyperplane transverse to
it. The orientation of the hyperplane, but not its extent, or
location in space-time is determined by the nonzero components of
$\bar{\psi} \gamma^{[\mu_1 \ldots \mu_k ]} \psi$ for all $k$.
 Equivalently, given a point in spacetime, the choice of a
pure spinor at that point can be thought of as determining the
direction and orientation of the holoscreen on which the
information at that point is encoded.   A hypothetical holoscreen
gauge invariance can thus be thought of as a local transformation
that changes a pure spinor at each point.   In $3,4,6$ and $10$
dimensions, pure spinors can be obtained as the projective space
defined by a linear representation of the Lorentz group (Dirac,
Weyl, Symplectic Majorana, and Majorana-Weyl).  In other
dimensions there is no canonical way to associate a general
spinor with unique pure spinors.  We can always find a basis for
the spinor representation consisting entirely of pure spinors.
Indeed, consider two null directions, one with positive and the
other with negative time component. The pure spinor conditions
for these two null vectors define two subspaces of the space of
all spinors, with half the dimension of the total space.
Furthermore, since a pure spinor uniquely determines its null
direction, the two subspaces are linearly independent, and thus
form a basis for the entire Dirac spinor space.  However, in
general this splitting is not canonical - the choices of null
vectors are arbitrary.  In $3,4,6,$ and $10$ dimensions, there is
a canonical way to decompose general spinors into irreducible
representations of the Lorentz group, such that each spinor in
the irrep is pure.

In any spacetime which is a compactification of a ten dimensional
theory, we can decompose a general spinor into pure spinors in a
unique way, so a general spinor at a point can be thought of as
defining the direction and orientation of a pair of holographic
screens on which to project the information from that point.  One
of the screens can be thought of as being in the past of the
point and the other in its future, according to the sign of
$p^0$. As a consequence, a gauge principle that refers to the
ambiguity of choosing a spinor at each point in spacetime (which
is to say, local SUSY) , can be thought of as the holographic
gauge invariance implicit in the work of Bousso.

One of the lacunae in my understanding of this subject is the
lack of an analogous statement about eleven dimensions.  I know
of no way to canonically split a general spinor into pure spinors
in eleven dimensions.  In the discussion below, this will be
dealt with in an unsatisfactory manner.  We will discuss eleven
dimensional Big Bang cosmology, where we will see that the
fundamental variables are past directed pure spinors.  This will
enable us to sidestep the necessity to decompose a general spinor
into pure spinors.

Aficionados of the superembedding formalism
\ref\sorokin{D.P.~Sorokin, {\it Superbranes and Superembeddings},
Phys. Rept. 329, 1, (2002), hep-th/9906142.} will undoubtedly
appreciate the deep connection between what I am trying to do
here and that formalism.  Superembedding is particularly powerful
in eleven dimensions.  I regret that my own understanding of this
formalism is so rudimentary that I cannot exploit its elegance in
the present work.

The projective invariance in the definition of pure spinors is a
classical gauge invariance.  It is real or complex, depending on
the nature of the Lorentz irreps in the given space-time
dimension.  When we quantize the pure spinors, we will break most
of this invariance, but there will always be at least a $Z_2$
gauge invariance left over in the quantum theory.   We will see
that this $Z_2$ can be identified with Fermi statistics.  Indeed,
the classical spinors of the CP equation are bosonic.  We will
quantize them as compact bosons - generalized spin operators.
Each such operator will correspond to a new bit of holographic
screen which is added to the Hilbert space describing the
interior of a backward light-cone in a Big Bang space-time, as
one progresses along a timelike trajectory.  Operators
corresponding to independent areas on the screen commute with
each other.  However, the $Z_2$ gauge invariance of the formalism
will enable us to perform a Klein transformation which makes all
the operators into Fermions.  In this new basis for the operator
algebra, the $Z_2$ is just $(-1)^F$.  In some dimensions there
will be a larger group of discrete gauge transformations, which
survives quantization of the CP equation.  I would like to
interpret this as a discrete R symmetry.

In the last section, I will use quantized CP spinors to construct
a holographic formulation of quantum cosmology.

\subsec{SUSY and Poincare: a marriage of convenience or
necessity?}

The arguments of the preceding subsections suggest strongly that
the asymptotically flat limit of a dS spacetime will be SUSic. It
is tempting to speculate that this is a special case of a more
general principle: quantum gravity in AF spacetimes {\it must} be
SUSic.   The temptation comes mostly from our abject failure to
find asymptotically flat SUSY violating vacua of perturbative
string theory.  Until recently, all attempts to violate SUSY lead
either to tachyonic instabilities, or the generation of a
potential for moduli.   In some of the latter situations (those
where the potential is positive and draws the system into the
semiclassical region of moduli space) , \ref\fsbd{T.~Banks,
M.~Dine, {\it Dark Energy in Perturbative String Cosmology}, JHEP
0110, 012, (2001), hep-th/0106276. } , it appears reasonable to
conjecture that there is a cosmological FRW spacetime with flat
or negatively curved spatial sections and only a Big Bang
singularity.  In cases where the potential is negative and draws
the system into a stongly coupled, string scale region, it is
likely that the AF perturbative vacuum has no resemblance to the
real physics. These models either do not really exist, or are
related to SUSY violating small radius AdS vacua.   More
generally, the failure to find an asymptotically flat SUSY
violating quantum state for gravity , is what is conventionally
called the cosmological constant problem. There are plausible
constructions of SUSY violating AdS vacua\evaetal\ , but they all
have the property that SUSY breaking vanishes as the AdS radius
is taken to infinity.  There is no known example of a SUSY
violating AdS vacuum in which SUSY violation appears to survive
the large radius limit.   Furthermore, all known SUSY violating
AdS vacua that can be constructed as concrete conformal field
theories have ``radius of order the string scale" in the sense
that they contain a plethora of operators which correspond to
particles of Compton wavelength of order the AdS radius.

In a beautiful paper, Kutasov and Seiberg \ks\ exhibited a clear
connection between SUSY violation and tachyonic instability in
perturbative string theory . A stringy Hagedorn spectrum
generally implies , via modular invariance, tachyonic instability,
which can only be removed by asymptotically exact SUSY
cancellations. One is tempted to speculate that the much more
rapidly diverging spectrum of black hole states in AF spacetime
might lead to instability if SUSY were not exact.

To attack this question, one must find a replacement for modular
invariance, which is a perturbative string theory concept.  I
would like to suggest that the replacement is simply crossing
symmetry, analyticity, and unitarity of the S-matrix.   The two
body scattering amplitude for spinless massless particles is an
analytic function of the kinematic invariants: $A(s,t)$. Here
$s>0$ is the square of the center of mass energy and $t<0$ the
invariant momentum transfer. Crossing symmetry is the claim that
when analytically continued to $s < 0$, $t > 0$ this is the same
amplitude (and a similar statement when the Mandelstam $u = - s -t
$ variable is the center of mass energy).

In the analysis of hadronic scattering amplitudes, it is well
known that the large $s$ fixed $t$ behavior of amplitudes is
dominated by singularities that probe the spectrum of the theory.
In particular, for large impact parameter we are probing the
lightest states in the theory, and would also encounter tachyons
if there were any.

In quantum gravity, asymptotic darkness implies that scattering
amplitudes at large $s$ and values of impact parameter that grow
like a power of $s$ are dominated by the production of black
holes. Elastic cross sections fall off exponentially in this
range of impact parameters because the probability that a black
hole formed in a two body collision will decay into two bodies is
Boltzmann suppressed . The two body final state has very little
entropy. The optical theorem must be satisfied by the
contribution of states at very large impact parameter.  The
Froissart bound violating growth of black hole cross sections is
enough to prove the existence of massless states in the theory,
and, at least at the level of naive Regge pole analysis, the fact
that there has to be a massless spin two particle.  Of course, we
already know that large impact parameter collisions are likely to
be dominated by (eikonal) graviton exchange.

I believe that a more careful analysis of the analyticity,
unitarity and angular momentum structure of high energy amplitudes
in the light of black hole dominance of high energy collisions,
might lead to an analog of the Kutasov-Seiberg result in
perturbative string theory. That is, analyticity, crossing and
unitarity of the S-matrix, plus black hole dominance of high
energy cross sections might generically imply a tachyon in the
crossed channel. Delicate supersymmetric cancellations might be
the only way to eliminate the tachyon. In an exactly SUSic
theory, any amplitude for production and decay of a black hole
will be exactly related to another amplitude in which an
additional soft gravitino is emitted, by a SUSY Ward identity.
The two amplitudes will behave differently under crossing
symmetry.

The idea that there are no Poincare invariant vacua that are not
Super-Poincare invariant is of course consistent with (though not
strictly implied by) the idea of cosmological SUSY breaking.  It
remains to be seen whether the almost forgotten lore of analytic
S-matrix theory will prove useful in addressing the problem of the
nature of quantum gravity and proving a deep connection between
SUSY and stability in quantum gravity.

\subsec{Discussion}

This section has been concerned with the role of SUSY in theories
of quantum gravity, and consists of three distinct claims. The
first (in a different order than I have presented them) is that
local SUSY is connected with the Holographic principle, and is
the local field theory limit of a gauge invariance corresponding
to the freedom of changing holographic screens.  The conventional
wisdom holds that non-supersymmetric theories of quantum gravity
might be sensible, while our claim is that every consistent
theory of quantum gravity is holographic and therefore a theory of
supergravity.

The second claim is that Poincare invariance occurs in theories
of quantum gravity only as a subgroup of super Poincare
invariance.  The evidence for this contention is our abject
failure to find a counterexample (which is ,in some sense, what we
usually call the cosmological constant problem).  I suggested a
more fundamental approach to this claim, based on examining the
compatibility of a black hole spectrum of states with the
existence of a unitary, Poincare invariant, analytic S-matrix.
Another approach to this question might be via the AdS/CFT
correspondence.  Unfortunately, we know too little about how the
full S-matrix is constructed in taking the large AdS radius limit,
to make much progress here.

I then presented a heuristic calculation of the scaling law
$m_{3/2} \sim \Lambda^{1/4}$ relating the gravitino mass to the
radius of de Sitter space.  This calculation makes TeV scale
superpartners a natural consequence of the (apparently) observed
value of the cosmological constant.   The calculation is only
consistent if the effective field theory explanation of SUSY
breaking is non-gravitational.  It also puts additional
constraints on the low energy theory, and implies in particular
that moduli must be frozen by SUSic dynamics at a high scale.
There is no cosmological moduli problem in a theory with
cosmological SUSY breaking.  There are other consequences for low
energy phenomenology.   Models where dark matter is a neutralino
are incompatible with this mechanism for SUSY breaking.  Finally,
a discrete R-symmetry is crucial to the mechanism.  It might be
related to a family symmetry for quarks and leptons.

There are two directions for further work on this mechanism.  The
first is to develop a more detailed mathematical model of the
quantum mechanics of a dS universe, which could lead to a more
rigorous derivation of the scaling law.  The second is to explore
the constraints on low energy dynamics more fully and come up
with an attractive, phenomenologically viable and predictive model
of TeV scale physics, based on these ideas.  Both directions are
being pursued.

\newsec{\bf Holographic cosmology}

In this section I will recall some ideas about holographic
cosmology that were presented in \ref\tbmill\
\ref\bfmcosmo{T.~Banks, W.~Fischler, {\it M-theory Observables
for Cosmological Space-Times}, hep-th/0102077.} , and extend them
by choosing the fundamental variables of the theory to be
quantized versions of the pure spinors of Cartan and Penrose. The
formulation of quantum cosmology that I will present is
Hamiltonian, and therefore necessarily gauge fixed.  Therefore,
the connection between CP spinors and a gauge principle for
holographic screens will not be immediately obvious.  However, we
will see that the fundamental variable of the theory is a path
dependent spinor.  I view it as the gauge fixed, holographic
version of the quantized gravitino field.

\subsec{Prolegomenon to a Holographic Theory of Space Time}

The treatment of cosmology in string theory has, for the most
part, been an exercise in effective field theory. Many
cosmological solutions of the equations of low energy
perturbative string theory can be found, but like most time
dependent solutions of Einstein's equations, they contain Big Bang
or Big Crunch singularities. This indicates the necessity for a
more profound approach to the problem. Fischler and Susskind
provided a fundamental new insight into cosmology, and Big Bang
singularities, by trying to impose the holographic principle in a
cosmological context\ref\lenwilly{W.~Fischler, L.~Susskind, {\it
Holography and Cosmology},hep-th/9806039. }. This work was
followed by Bousso's construction of a completely covariant
holographic entropy bound\raph\ .

I believe that the work of F(ischler) S(usskind) and B(ousso)
provided us with the foundations of a quantum theory of cosmology.
There are three important principles that are implicit in the work
of FSB:

\noindent 1. The holographic principle is consistent with the idea
of a {\it particle horizon}, a notion which we generally derive
from local field theory.  More generally, it is consistent with
the idea that a {\it causal diamond} in spacetime contains an
operator algebra that describes all measurements, which can be
performed within this diamond.  In many cases, the holographic
principle implies that the dimension of this operator algebra is
finite. In this case there is a unique Hilbert space
representation of the algebra.

\noindent 2. In particular, the interiors of backward light cones
in a Big Bang spacetime must have finite operator algebras.
Furthermore, the FSB entropy bound implies that the dimension
decreases (apparently to zero) as we go back to the Big Bang
singularity. These results lead one to conjecture that instead,
the (reverse) evolution stops when the Hilbert space has some
minimal dimension.  They also lead one to some guesses about the
fundamental formulation of quantum cosmology, which I will sketch
below.

\noindent 3.  Within the semiclassical approximation the
holographic principle is compatible with
F(riedman)-R(obertson)-W(alker) cosmology at early times if and
only if the stress tensor satisfies the equation of state
$p=\rho$, with the entropy density related to the energy density
by $\sigma \propto \rho^{1\over 2}$ \foot{The latter condition
means that the homogeneous modes of minimally coupled scalars
{\it do not} satisfies the requirements of holography.}. This is
a peculiar new form of matter.  Fischler and I
\ref\holocosmo{T.~Banks, W.~Fischler, {\it An Holographic
Cosmology}, hep-th/0111142.} gave a heuristic picture of such a
system as a \lq\lq dense fluid of black holes \rq\rq , but a more
precise quantum description still eludes us.

I would like to sketch the outlines of a quantum cosmology based
on these principles.  This sketch is an update of ideas presented
in \tbmill\ref\bfmcosmo{T.~Banks, W.~Fischler, {\it M-theory
Observables for Cosmological Space Times}, hep-th/0102077.} . It
is still less than a full dynamical quantum theory of spacetime.
In presenting it, I have used the following strategy. I utilize
spacetime concepts to motivate quantum mechanical constructions.
Eventually, one would like to turn everything around, and present
a set of purely quantum axioms from which we derive a classical
spacetime.  The reader should keep in mind the dual purpose of
this discussion and, as it were, try to read every argument both
backwards and forwards .

Consider a timelike trajectory (perhaps a geodesic) in a Big Bang
spacetime, and a sequence of backward light cones whose tips end
on this trajectory.  The FSB bound implies that the Hilbert space
describing all measurements in the interior of each of these light
cones is finite dimensional. Let us define the entropy to be the
logarithm of this dimension (it is the entropy of the maximally
uncertain density matrix on this Hilbert space). Let us for the
moment restrict attention to a period in which the universe is
expanding. Then the entropy decreases as we follow the trajectory
back to the Big Bang.

The concept of particle horizon means that each of these Hilbert
spaces should have a self contained description of all of the
physics that goes on inside it.  That is, there should be a
sequence of unitary transformations describing time evolution
inside each backward light cone, without reference to any of the
larger light cones.  On the other hand, the dynamics in a large
light cone should be restricted by consistency with earlier light
cones in the sequence.  The reason that I insist on a sequence of
unitaries, rather than a continuous one parameter family (a
groupoid\foot{I would like to thank G. Moore for explaining the
mathematical name for the composition property for time evolution
operators in time dependent quantum mechanics.}) is that the
system in any one light cone is finite dimensional.  A finite
system can have a continuous time evolution if it is in contact
with an external classical measuring apparatus, but, because of
the time-energy uncertainty relations (whose precise form depends
on the spectrum of the finite system) it does not make sense to
talk about infinitely precise time resolution as a measurement
performed by the system on itself. A more fundamental reason for
discreteness will be discussed below.

This suggests, that as time goes on and the particle horizon
expands, more and more precise time resolution becomes available.
Thus, the time intervals between unitary transformations in the
sequence should not be thought of as defining the Planck time.
Instead, I insist that they define time slices in which the FSB
area increases by some minimal amount (to be quantified below).
Call the sequence of Hilbert spaces ${\cal H}_n$. $\hn$ has
dimension $K^n$, and we think of it as defining the Hilbert space
inside a backward light cone whose holographic screen has FSB area
$4 n {\rm ln} K$ in Planck units. In each $\hn$ there is a
sequence of unitary transformations $U_n (k)$ for $n \geq k \geq
0$. One further assumes that $\hn = H \otimes {\cal H}_{n-1}$,
where $H$ is a $K$ dimensional space.  The maps $U_n (k)$ are
required to factorize in a manner compatible with this
concatenated tensor factorization of the Hilbert space.  For
example, for every $n$ and $k$, $U_n (k)$ for $k < n$ is a tensor
product of $U_{n-1} (k)$ and a $K$ dimensional unitary
transformation on $H$.  Below, we will choose the number $K$ in a
natural way that depends on the dimension of spacetime.

This definition gives us some idea of how much time is represented
by each unitary evolution in the sequence.  An area $4n\ {\rm ln}\
K$ in $d$ spacetime dimensions, allows the creation of black holes
of energy of order $(4n\ {\rm ln}\ K)^{(d-3)\over (d-2)} $.  The
inverse of this energy is the maximum time resolution that such a
system can have.   On the other hand, if we make some other
assumption about the state of the system, we may have less time
resolution than this.  Thus, we can begin to see a correlation
between the spacetime geometry and the matter content of the
system.

We also see the fundamental reason for discreteness in these
equations.  The FSB areas of backward light cones in a Big Bang
space-time are quantized because they refer to the logarithms of
Hilbert space dimensions.

 So far of course we have defined much less than a full
spacetime. To go on, we need to consider neighboring timelike
trajectories, and we must introduce the dimension of spacetime.
To do this, introduce a $d$ dimensional cubic lattice, and assign
Hilbert spaces and unitary operators to each vertex of the
lattice.

There are several disturbing things about this (as far as I can
see) unavoidable introduction of dimensions.  The is that we
believe that we can define cosmologies in string theory, that
interpolate between spaces of different dimension. For example,
the Kasner cosmologies studied in \ref\bmfbm{ T.~Banks,
W.~Fischler, L.~Motl, {\it Duality vs. Singularities}, JHEP 9901
(1999) 019, hep-th/9811194; T.~Banks, L.~Motl, {\it On the
Hyperbolic Structure of Moduli Space With Sixteen SUSYs}, JHEP
9905:015,1999, JHEP 9905:015,1999.} can interpolate between
heterotic strings on tori and 11D SUGRA on K3 manifolds. It is
not clear to me that this is a difficulty.  We are not describing
local field theories here, and our description might be valid in
all regions of moduli space, even though defined with respect to
one.  What is certain, is that all of these dualities involve the
nontrivial topology of the compactification manifold. We can for
the moment restrict our attention to describing the noncompact
part of space, with the compact parts described by the structure
of the spectrum of states in Hilbert space.  However, there is
obviously much to be understood about this question. In the
present paper I will restrict attention to non-compact $11$
dimensional cosmologies.

The second disturbing aspect of our construction will be an
asymmetry between space and time. It is intrinsic to our
formulation of the problem in terms of time evolution in Hilbert
space (rather than some sort of path integral formalism). We have
chosen a rather particular gauge, in which every point on a time
slice has a backward lightcone with equal FSB area.  One could
make different choices, but none would be gauge independent.  No
physical Hamiltonian of a general covariant theory can be gauge
independent, since the choice of time evolution is a choice of
gauge.  Only in spacetimes with a fixed classical asymptotic
boundary can we imagine a gauge independent choice of Hamiltonian.
We will introduce the asymmetry between space and time into our
notation by labeling points in the lattice by a $d$ vector of
integers $(t, {\bf x})$.

Now we have to address the question of how the Hilbert spaces and
time evolution operators corresponding to different points on the
lattice, are related to each other.  It is here that the formalism
parts company with a lattice field theory like system, where each
point should have independent degrees of freedom.  In fact, since
we are associating the observables with experiments done in the
backward light cone of the point, there should be a large degree
of overlap between nearest neighbors. Indeed, we defined the
smallest time difference by insisting that the Hilbert space at
time $n$ have only $K$ times as many states as that at time
$n-1$.  If $K$ were $2$, this would be the minimum increase
compatible with the notion that the new particle horizon has some
independent degrees of freedom in it that were not measurable in
the old one. Similarly we will require a maximal overlap for
nearest neighbor points on the lattice.  That is, the Hilbert
spaces $\hn (x) $ and $\hn (x + \mu)$ should each factorize as
\eqn\factor{\hn (x) = H(x) \otimes O(x,x+\mu )\ \ \hn (x+\mu ) =
H(x+\mu ) \otimes O(x, x+\mu ), }  where for each $y$, $H(y)$ is a
$K$ dimensional space.

We will choose $K$ in a manner motivated by our remarks about the
connection between supersymmetry and holography.  Let $S_{\alpha}$
transform in the irreducible spinor representation of the Lorentz
group $Spin(1,d-1)$.  The details of the construction will depend
somewhat on the properties of spinors in various dimensions, so I
will restrict attention to d=11.   We will insist that the spinor
be pure, that is , that $\bar{S} \gamma^{\mu} S \gamma_{\mu} S =
0$. Such spinors have $16$ independent real components.   In the
quantum theory, they will be quantum operators, $S_a$, $a=1\ldots
16$ .  We also restrict attention to past directed pure spinors -
the associated null vector is past directed.  In choosing to
describe the pure spinor in terms of only sixteen variables, we
have chosen a gauge for local Lorentz gauge symmetry.  In
principle, one could keep $32$ components and a local symmetry
which allowed us to reduce to $16$.  However, the Lorentz
connection would have to be a constrained variable, in order not
to introduce new degrees of freedom into the system.  We are
aiming toward a completely gauge fixed Hamiltonian description of
our cosmology. Below, we will introduce a mapping $\Psi$ between
the operator algebras in Hilbert spaces at different points on the
lattice.  In particular, that mapping will relate the spinor
basis at one point to that at another. $\Psi$ implicitly contains
the gauge fixed Lorentz connection.

We have seen that, classically, a past directed pure spinor
determines a past directed null direction.  We think of the
physical interpretation of this null direction in terms of two
holographic screens for an observer traveling along the timelike
trajectory between $(t,{\bf x})$ and $(t+1, {\bf x})$ .   The
physics inside the backward light cone of the observer at these
two points, can be projected onto a pair of holographic screens,
both in the past of the tips of the light cones. In a geometrical
picture, the information that is not contained in the smaller
screen can be communicated to the observer at some point, P,  on
his trajectory between the tips of the two light cones. The new
pure spinor that we add to the system may be thought of as the
instruction for building the new piece of the holographic screen,
on which the information at P is to be projected.  Of course,
this classical language can have only a poetic meaning at the
time scales on which we are making our construction.

It is important to note that the paragraph above contains the
answer to the question of where and how large the holographic
screen is.  If we assume that the quantum formalism will, in the
limit of large Hilbert spaces, indeed determine a classical
geometry consistent with the words we have been using, then the
bit of holographic screen that is added by the operator $S_a
(t,{\bf x})$ is located on the FSB surface of the backward
light-cone from $(t+1, {\bf x})$ , and has area $4\ {\rm ln}\
256$ in Planck units.  The null vector which would specify
precisely where on that screen this particular variable is, is
the bilinear current constructed from this pure spinor.  It is a
quantum operator, and so only describes probability amplitudes
for the bit of screen to be at specific points on the FSB surface.
The FSB surface itself is constructed out of all the spinors in
the Hilbert space ${\cal H}_{n+1}$, so its quantum fluctuations
are small in the limit of large area.

As anticipated above, we will build the Hilbert space ${\cal H}(t
+ 1, {\bf x}) $ by adding operators $\hat{S}_a (t+1, {\bf x})$ to
the Hilbert space ${\cal H}(t, {\bf x})$. These will commute with
all of the operators in the latter space.   The defining relation
for a pure eleven dimensional spinor is invariant under real
projective transformations of the spinor. We will break this
invariance in the quantum theory.

Thus, we postulate that \eqn\acr{[\hat{S}_a , \hat{S}_b ]_+ =
2\delta_{ab}.}  Up to normalization, this is the unique ansatz
that gives a finite dimensional Hilbert space, and is invariant
under the $SO(9)$ group of rotations that leave the null vector
invariant.  These postulates break the projective invariance
except for a factor of $(- 1)$.  We will treat the latter factor
as a $Z_2$ gauge transformation, which will eventually be seen as
Fermi statistics.  The fact that the classical projective gauge
symmetry of the CP equation is broken down to $Z_2$ has to do with
the fact that our spinor carries information about the conformal
factor of the spacetime geometry, as well as its causal
structure.  Indeed, the commutation relations determine the
dimension of the new Hilbert space, and thus the area of the new
holographic screen.  The logarithm of the dimension of the new
Hilbert space increases by $8\ {\rm ln}\ 2$, which corresponds to
an increase in area (Planck units) of $32\ {\rm ln}\ 2$.

We can now turn the $\hat{S}_a$ into Fermions, by defining $S_a =
(-1)^F \hat{S}_a$, where, $(-1)^F$ is the product of all of the
previous $S_a$ operators (note that the number of these operators
is always even) . In other words, we start with the irreducible
representation of the Clifford algebra. This defines the smallest
possible Hilbert space at the moment of the Big Bang. Then we
build successive Hilbert spaces along a given timelike
trajectory, by tensoring in one more commuting copy of the
minimal Clifford representation. We then do a Klein
transformation to present the full algebra as a larger Clifford
algebra.  The Klein transformation is a $Z_2$ gauge
transformation, which is the quantum remnant of the projective
invariance of the Cartan-Penrose equation.   It is Fermi
statistics of the Klein transformed operators.  Note that all
operators transforming in integer spin representations of the
Lorentz group, will be even functions of the $S_a$, so the
connection between spin and statistics is built into the
formalism.

To recapitulate, the quantum description of the causal pasts of a
sequence of points along a given timelike trajectory in a Big
Bang cosmology is described by a sequence of Hilbert spaces
${\cal H}_n$.   The operator algebra of the $k$th Hilbert space
is the Clifford algebra generated by operators $S_a (n)$ with $1
\leq n \leq k$:

 \eqn\onehor{[S_a (n), S_b (m)]_+ = \delta_{ab}\delta_{mn}}
The operator $S_a (n)$ in each Hilbert space may be identified
with the operator with the same labels in any other Hilbert space.
We will see later that this identification may be viewed as a
gauge choice for the discrete analog of local SUSY.

Dynamics is defined by a sequence of unitary transformations, $\{
U(n) \}$, in each Hilbert space, satisfying a simple compatibility
condition, which will be discussed below.  In principle, we could
introduce a continuous unitary groupoid $U(t,t_0 )$ such that the
unitary transformations in the sequence could be viewed as the
values of $U(t_n, 0 )$ at a sequence of times.  In this way the
formalism becomes that of ordinary quantum mechanics, with a time
dependent Hamiltonian,  but changes in the groupoid, which do not
change the values at the special times $t_n$, should be viewed as
physically equivalent.

The sequence in ${\cal H}_k$ has $k$ steps, and should be thought
of as the evolution operators over the time steps determined by
the sequence of points on our timelike trajectory.  The
fundamental consistency condition is that the operator $U_k (n)$
in ${\cal H}_k $with $n < k$ should be a tensor product of the
($k$th copy of) $U_n (n)$ with an operator that depends only on
the $\hat{S}_a (m)$ with $ m > n$.   Thus \eqn\consist{ U_k (n) =
U_n (n) V_k (n),} where $V_k (n)$ is a function only of $S_a (m)$
with $m >n$.  We will impose the $Z_2$ gauge invariance on all of
these unitaries, so that they are even functions of the
fundamental variables, and we can ignore the distinction between
the hatted and bare headed variables.  Note also, that in writing
the last equation we have used the same notation for the operator
$U_n (n)$ and the copy of this operator in every ${\cal H}_k $
with $k > n$.

These rules define a quantum system, which is compatible with the
notion of particle horizon in a Big Bang cosmology.  The Hilbert
space ${\cal H}_k$ describes all measurements that can be done
inside the particle horizon at time $t_k$, in a manner compatible
with the fact that measurements inside earlier particle horizons
commute with measurements that can only be made at later times.
Each particle horizon has its own time evolution operator, but
the evolution operators at early times, agree with those in
previous particle horizons, in their action on those variables
that are shared between the two systems\foot{From here on I will
stop insisting that the shared operators are really copies of the
operators at earlier times. The reader will have to supply this
pedantry by himself.  Its importance will be apparent when we
discuss copies of operators associated with other timelike
trajectories.}

The system is also compatible with the holographic principle in
that we will identify the dimension of the Hilbert space with the
area of the FSB surface on the past light cone\foot{This
identification is only appropriate for the past light cones on
trajectories in an eternally expanding universe. In contracting
universes, the FSB area can sometimes decrease as one goes into
the future. It must then be interpreted as the entropy of a
density matrix more pure than the uniform probability density. The
interpretation of this is that the assumed spacetime geometry and
matter content is a very special class of states of the system.
More generic initial conditions at times before the FSB area
began to decrease would have led to a different spacetime in
these regions.  Of course, the latter statement could also be
made about, {\it e.g.}, the future evolution of an expanding
matter dominated FRW universe. However, in this case we can still
imagine exciting a more general configuration in the future by
creating lots of black holes. In contracting regions, certain
possible excitations of the system at early times are ruled out
by the assumption that the geometry behaves in a particular
classical manner.  In this connection note that if we examine
contracting FRW universes with matter with equation of state
$p=\rho$, corresponding to a maximal entropy black hole fluid,
then the FSB area of backward light cones always increases.  It
is only the assumption that low entropy systems with soft
equations of state persist into contracting regions that leads to
the phenomenon of decreasing area. }. This statement does not have
much content until we enrich our system and show that it does
have a spacetime interpretation.

Indeed, the conditions we have stated so far are very easy to
satisfy, and most solutions do not resemble spacetime in any
obvious way.   What is missing is the notion that the new degrees
of freedom that come into a particle horizon ``come from other
points in space".   To implement this, we return to our hypercubic
eleven dimensional lattice with points labeled $(t, {\bf x})$ and
Hilbert spaces ${\cal H} (t, {\bf x})$.  The sequence of Hilbert
spaces at fixed ${\bf x}$ has all the properties we have
described above.

To understand the geometric interpretation of , {\it e.g.}, the
Hilbert space ${\cal H} (t, {\bf x + e_1}$, where ${\bf e_1}$ is
some unit lattice vector, introduce a time slicing of our Big
Bang spacetime by the rule that the past light cone of every
point on a time slice has equal FSB area (for FRW these are just
slices of cosmic time, but a single unit in $t$ does not
correspond to a fixed unit of cosmic time, but rather a fixed
unit of FSB area) .  Now, starting at a point labeled ${\bf x}$
on a fixed time slice, choose a spacelike direction on the slice
and move along it to a new point, labeled ${\bf x + e_1}$.  The
intersection of the past light cones of these two points has an
almost everywhere null boundary but is not a full light cone.
Choose the point (for very small distance between the two points
on the time slice, it will be unique) inside the intersection
whose past light cone has the largest FSB area, and call this the
FSB area of the intersection . The point ${\bf x} + {\bf e_1}$ is
chosen such that the FSB area of the intersection is smaller than
the FSB areas of the causal pasts of ${\bf x}$ and ${\bf x} +
{\bf e_1}$, by precisely the fundamental unit. Now proceed to do
the same in the negative $e_1$ direction and in $9$ other locally
independent directions. Then repeat the same procedure for each
of these new points and so on {\it ad infinitum} (for this paper,
we restrict attention to spatial topologies, which are trivial and
extend to infinity in all dimensions).   Repeat the same for each
time slice.  This picture motivates our lattice of Hilbert spaces.

The crucial step now is to introduce maps between a tensor factor
of the operator algebra (equivalently, the Hilbert space, since
everything is finite dimensional) in ${\cal H} (t, {\bf x}) $ and
that in ${\cal H} (t, {\bf x + e_i})$ for every (positive and
negative) direction.  The common factor Hilbert space has
dimension smaller by a factor of ${1\over 256}$.  Equivalently we
can think of this as a relation, which defines a copy of the
generators of the algebra at ${\bf x}$ in the Hilbert space at
${\bf x + e_i}$.

\eqn\connection{S_a (t_i; t , {\bf x}; {\bf x + e_1}) = \Psi_{ab}
(t_i , t_j; {\bf x, x+e_1}) S_b (t_j , {bf x + e_1})}

The labels $0\leq t_i \leq t ,0\leq t_j \leq t$ on the operators,
remind us that the Hilbert space ${\cal H} (t, {\bf x})$ contains
operators that have been copied from the Hilbert spaces at all
previous times.  The map $\Psi$ is part of the definition of the
dynamics of this quantum spacetime.   It is subject to a large
number of constraints.  Viewed as a matrix on the $256 t$
dimensional space of $S$ components, it should have rank $256
(t-1)$.  Precisely $256$ generators of the algebra at ${\bf x}$
should have vanishing representative in the nearest neighbor
Hilbert space. Furthermore, the different $\Psi$ maps at different
points of the spacetime lattice must all be compatible with each
other.

A much stronger set of constraints comes from requiring that the
unitary transformations $U(t_k, 0)$ in each Hilbert space be
compatible with each other after application of the map $\Psi$.
This is a system of mutual compatibility constraints between the
$\Psi$ maps and the unitary transformations. Indeed, one is
tempted to conjecture that any lattice of Hilbert spaces, $\Psi$
maps and unitary transformations satisfying all of these axioms
should be viewed as a consistent quantum mechanical description
of a Big Bang cosmology.  I am not prepared to make such a bold
conjecture at this time.  Many examples will have to be
discovered and worked out before we can hope to understand this
formalism, and whether it needs to be supplemented with
additional axioms.

The bilateral relations between nearest neighbor Hilbert spaces
on the lattice, enable us construct copies of subalgebras of the
operators in any Hilbert space, inside the operator algebra of
any other. For a pair of points on the lattice, this
correspondence will be path dependent.   Thus, in some sense, the
fundamental dynamical variables in the theory are the path
dependent objects

\eqn\gravitino{S^{\Gamma}_a (t, {\bf x} ; t^{\prime}, {\bf
x^{\prime}}).}

In words, this is the copy of $S_a (t, {\bf x})$, in ${\cal H}
(t^{\prime} , {\bf x^{\prime}} ) $ obtained by concatenating the
$\Psi$ maps along the path $\Gamma$ between the two points. The
$\Psi$ map gives us a special case of these variables for the
minimal path between nearest neighbor points. Thus, $S_a (t, {\bf
x} ; t, {\bf x + e_1}) $ can be thought of as a discrete analog
of the gravitino field $\psi_{a \mu} de_1^{\mu }$ integrated
along the link between two nearest neighbor lattice points.

We now see that the simple mapping between the operator algebras
at different times, at the same spatial point, can be viewed as a
gauge choice for the time component of the gravitino field. We do
not yet have any evidence that this formalism reduces to some
kind of classical field theory in limiting situations, but it
seems like that if it does, that field theory will be locally
SUSic.

\subsec{Discussion}

The system that we have been discussing bears some resemblance to
a lattice quantum field theory.  This is both misleading, and
suggestive.  It is misleading because the fields at different
points of the lattice at the same time do not (anti)-commute with
each other. Their commutation relations are complicated and
depend on the choice of $\Psi$ mappings. This choice is part of
the specification of the dynamics of the system. Note further
that if the true connection between geometry and quantum
mechanics is to be extracted from the entropy/area relation, the
space-time geometry will not be that of the lattice.  The lattice
does serve to specify the topology of the spacetime.

The relation to lattice field theory is however suggestive of the
possibility that in some dynamical circumstances, suitable
subsets of the variables of this system might behave like quantum
fields.

\subsec{ What is To Be Done?}

The answer to this question is of course: ``Almost everything".
More specifically, the most urgent problem is to find one example
of a solution to the constraints postulated above, and show that
in the large time limit it has an approximate description in
terms of quantum fields in curved spacetime.   The obvious case
to start with is that of a homogeneous isotropic universe.  That
is, every sequence of Hilbert spaces ${\cal H} ({\bf x}, t)$ has
the same set of unitary maps $U(s,s-1 )$ for $s \leq t$.  There is
essentially a single $\Psi$ mapping, which must be consistent with
the unitary dynamics.  This problem is still complicated enough
that no solutions have been found as yet.

With W. Fischler, I have conjectured a possible solution,
corresponding to a homogeneous spatially flat universe with
equation of state $p=\rho$.   Write $U(s,s-1) = e^{H(s)}$ and
expand $H(s)$ in powers of the Fermion operators.  In particular,
there will be a quadratic term

\eqn\hquad{H_2 (s) = \sum_{p,r < s} S_a (p) h(s|p,r) S_a (r)}

Now, for each $s$, let $h(s | p, r)$ be a random antisymmetric
matrix, chosen from the Gaussian ensemble.  It is well
known\ref\kwd{F.J.~Dyson, J. Math. Phys. 3, 140 (1962), 3, 1191,
(1962), 3, 1199, (1962) ; V.~Kaplunovsky, M.~Weinstein, {\it
Space-Time: Arena or Illusion?}, Phys. Rev. D31, 1879, (1985),
use this connection in a manner similar to what is presented
here.} that large Gaussian random antisymmetric matrices have a
spectral density that behaves linearly in a larger and larger
region around zero eigenvalues.  Thus, for large $s$, the
spectrum of the time dependent Hamiltonian $H_2 (s)$has a
universal behavior that looks like that of a system of free
massless Fermions in $1+1$ dimensions.  Furthermore, with the
exception of a single marginally relevant four Fermion operator
(the analog of the Bardeen-Cooper-Schrieffer operator), this low
energy spectrum will not be disturbed by higher order polynomials
in Fermions. The universal behavior of the spectral density is
shared by a large class of random Hamiltonians for the Fermion
system.

Thus, for large $s$, although we are dealing with a problem with
time dependent Hamiltonian, we approach a system whose spectral
density becomes time independent and satisfies the energy/entropy
relation $\sigma \sim \sqrt{\rho}$ of a $p=\rho$ fluid.  Note
that, in the hypothetical translation of this physics into a
spacetime picture, the energy of this system at time $t$ would be
interpreted as the energy density at the tip of the backward light
cone, $(t, {\bf x})$.

On the other hand, because for each $s$ we make an independent
choice of random Hamiltonian, there is no sense in which the
quantum state of the system settles down to the ground state of
any given Hamiltonian, even after time averaging.   All of the
degrees of freedom of the system remain permanently excited. The
density matrix of the system is completely random, maximizing the
entropy, but certain energetic properties become smooth and
universal for large $s$.  To prove that this system satisfies our
axioms, one would have to exhibit a $\Psi$ mapping compatible
with this prescription for time evolution. This has not yet been
done.

\subsec{Discussion}

 Our discussion has been restricted to
Big Bang cosmological spacetimes.  I have emphasized above and
elsewhere\ref\hetero{T.~Banks, {\it A Critique of Pure String
Theory}, Talk given at Strings 2002, Cambridge, UK.} that one
should expect the fundamental formulation of quantum gravity to
depend on the asymptotic geometry of spacetime. Gravity is not a
local theory, once one goes beyond the realm of classical geometry
(where the degree of non-locality can be controlled by the choice
of initial conditions of the classical solution), and its
fundamental formulation has every right to depend on the boundary
conditions .  Nonetheless, the considerations of the present
paper suggest the possibility of a more local, but perhaps gauge
dependent formulation, as has been advocated by Susskind. Again,
the idea is that the choice of holographic screen is a gauge
artifact. In asymptotically flat or AdS spacetimes, it may be
convenient and elegant to place the screen at infinity, but there
may also be other gauges where the same information is mapped
onto a collection of local screens. Our formulation of quantum
cosmology has fixed a particular gauge defined by equal area time
slices.

In asymptotically flat spacetime, the causal diamond formed by
the intersection of the causal past of a point with the causal
future of point in that causal past, has finite FSB area. Thus,
one can imagine assigning finite dimensional operator algebras to
causal diamonds and trying to imitate the formalism of this paper.
Any given finite dimensional algebra would be embedded in a
sequence of algebras corresponding to larger and larger causal
diamonds.  The limiting Hilbert space would be infinite and the
limiting time evolution operator in this space would approach the
scattering matrix.   Again, the formalism would be constrained by
the requirement of consistency with many partially overlapping
sequences corresponding to nested causal diamonds centered around
different points in space.

In asymptotically flat spacetime we expect to have an exact
rotational symmetry.     Thus, it makes sense to choose
holographic screens which are spherically symmetric. One would
want to represent the pure spinor operators for a finite causal
diamond as something like elements of a spinor bundle over a
fuzzy sphere in order to have a formalism which preserves
rotational invariance at every step.   Furthermore, the explicit
breaking of TCP invariance which was evident in our treatment of
Big Bang cosmologies, should be abandoned.  Each causal diamond
of FSB area $K^N$ should have a sequence of unitary operators
$U(t_k , - t_k)$ for $ 1 \leq t_k \leq N$, which commute with an
anti-unitary TCP operator.  In the limit as $N\rightarrow\infty$,
$U(t_N , - t_N)$ would become the S-matrix.  An important aspect
of this limit is that the finite $N$ fuzzy sphere should become a
conformal sphere as $N\rightarrow\infty$, in order to obtain a
Lorentz invariant S-matrix (the Lorentz group is realized as the
conformal group of null-infinity).

In asymptotically AdS spacetime, things are more complicated. The
causal past of a point includes all of AdS space prior to some
spacelike slice.  Thus if we try to construct the causal diamond
corresponding to a pair of timelike separated points, it becomes
infinite when the timelike separation is of order the AdS radius,
and the backward and forward light-cones of the two points
intersect the boundary of AdS space before intersecting each
other. This suggests that there should be a sequence of finite
dimensional operator algebras which cuts off at some finite
dimension of order $e^{R_{AdS}^{(d-2)}}$.   Since we already have
a ``complete" formulation of the quantum theory of AdS spacetimes,
it would seem to be a good strategy to search for such a sequence
of nested operator algebras within the Hilbert space of conformal
field theory. This would be a new approach to the puzzle of how
local data is encoded in the CFT.

\newsec{\bf Conclusions}

It is unfortunate but perhaps inevitable that the negative
conclusions of a paper like this are on a firmer footing than the
attempts to make progress in new and positive directions.  Unless
one rejects the AdS/CFT prescription for quantum gravity in Anti
de Sitter space, it is difficult to defend the idea that there is
a unique theory of quantum gravity, with different realizations
of it corresponding to minima of an effective potential.  This
field theory inspired picture is based on a separation between UV
and IR physics which is simply not there in theories of quantum
gravity.   I  have tried to investigate both real \isovac\ and
virtual \heretic\ transitions between vacua with different values
of the cosmological constant, or isolated vacua with the same
values of the cosmological constant and found that they do not
occur - black holes get in the way.  There remains one question
in this general category whose answer remains unclear: is there a
meaning to meta-stable dS minima which can decay into negatively
curved FRW spacetimes with vanishing cosmological constant.  It
would seem that the primary challenge here would be to establish
the existence of a quantum theory corresponding to the Big Bang
FRW universe, into which the metastable dS states are supposed to
decay.  Only with a well controlled quantum theory of this Big
Bang spacetime in hand, could we hope to make a rigorous
verification of the existence of dS ``resonances".   Calculations
of approximate effective potentials without an underlying high
energy theory, cannot resolve this question.

These results, to my mind, establish the existence of a variety
of consistent mathematical models of quantum gravity and lead to
the question of what distinguishes our world from among them.  I
believe that the key question to ask here is ``What is
Supersymmetry and Why Don't We See It?".

I have given a variety of partial answers to this question, of
varying degrees of plausibility.   At the deepest level, I
suggested that local SUSY was connected to the holographic
principle via the Cartan-Penrose equation.  I have a strong
feeling that there is something right about this idea, and an
even stronger one that I have as yet expressed it only clumsily.

The question of global SUSY depends, as does everything in
theories of quantum gravity, on asymptotic boundary conditions in
space-time.  In AdS space-time, global SUSY seems to be necessary
in order to obtain an AdS radius large compared to the string
scale. SUSY violating AdS theories with radius larger than the
string scale have been exhibited using low energy effective
Lagrangians, and flux compactification.  There is as yet no
proposed CFT dual for these models.  At any rate, the amount of
SUSY violation vanishes as the AdS radius goes to infinity.
Consistent with this, we have no example of a SUSY violating
theory of quantum gravity in asymptotically flat space-times. I
have conjectured that none exists.  The two avenues by which one
might try to establish or falsify this conjecture are to search
for SUSY violating sequences of CFT's with AdS radius going to
infinity in string units, or to use the combination of unitarity,
crossing, analyticity and black hole dominance of high energy
processes to prove the necessity of SUSY.

The {\it Poincare $\rightarrow$ Super-Poincare} conjecture leads
to, but is stronger than, the CSB conjecture that SUSY breaking in
the world we see is connected to the positive value of the
cosmological constant.   I have given an argument for the
anomalous scaling $m_{3/2} \sim \Lambda^{1/4}$, but to make it
rigorous one really needs a complete mathematical quantum theory
of dS space.  Absent such a theory, there is still an interesting
calculation which can be done to make this speculation more
plausible.  I have argued that the global coordinate picture of
an infinitely expanding sphere may actually be sensible in the
field theory approximation, if we make an infrared cutoff when
the volume of the sphere exceeds $R_{ds}^{1\over 2} $ in Planck
units.  Perturbative quantum gravity calculations in dS space are
fraught with IR divergences.  In particular, one might imagine
that the gravitino mass term in the low energy effective
Lagrangian might have (probably logarithmic) IR divergent loop
corrections.  Since we have argued that the IR cut-off is the dS
radius, this would suggest an anomalous dependence of the
gravitino mass on the cosmological constant.  Perhaps one could
even get the right critical exponent by adroit resummation of the
perturbation series.

I will end by mentioning two other directions of research that
are suggested by CSB.  The first is the program begun in
\scpheno\ to find a low energy description of the SUSY breaking
mechanism.  One must find a SUSic, R-symmetric theory, which,
when perturbed by R violating terms which are a function of the
cosmological constant, breaks SUSY and gives a gravitino mass of
order $\Lambda^{1/4}$.  The second is the search for an algorithm
which describes all ${\cal N} = 1, d=4$ SUSic compactifications
of M-theory, including isolated ones.   Arguments of holomorphy
suggest that the otherwise oxymoronic program of ``computing the
superpotential on moduli space", and finding stationary points
where it vanishes is a good first attempt at such an algorithm.
It would be of great interest to find a non-perturbative
formulation of this problem, perhaps as a sort of topological
version of M-Theory.

\listrefs

\bye